\documentstyle[prb,aps,amsfonts,amssymb,twocolumn,epsfig]{revtex}
\begin{document}
\draft
\sloppy

\newcommand{\w}{\omega}
\newcommand{\kf}{k_F}
\newcommand{\kappaV}{{{\mbox{\boldmath $\kappa$}}}}
\newcommand{\surfint}{\int \!\!\!\! \int}
\newcommand{\Q}{{\bf Q}}
\newcommand{\k}{{\vec{k}}}
\newcommand{\q}{{ \vec{q}}}
\newcommand{\A}{\text{\AA}}
\newcommand{\IM}{\text{Im}}
\renewcommand{\vec}[1]{{\bf #1}}
\wideabs{
\title{Magnetotransport in
nearly antiferromagnetic metals}
\author{A. Rosch}
\address{Center for Materials Theory, Department
of Physics and Astronomy, Rutgers University, Piscataway, NJ 08854, USA}
\date{\today}
\maketitle

\begin{abstract}
We present a theory of the magnetotransport in weakly disordered
metals close to an antiferromagnetic quantum-critical point. The
anisotropic scattering from critical spin fluctuations is strongly
influenced by weak but isotropic scattering from small amounts of
disorder.  This leads to a large regime where the resistivity obeys a
scaling form $\rho=\rho_0+\Delta \rho \approx \rho_0+T^{3/2} f(
T/\rho_0,(p-p_c)/\rho_0,B/\rho_0^{3/2})$, where $\rho_0$ is the
residual resistivity, $B$ the magnetic field and
$p-p_c>0$ measures the distance from the quantum-critical point on the
paramagnetic side of the phase diagram. Orbital effects of the
magnetic field are most pronounced in very clean samples for not too
low temperatures, where the resistivity for increasing magnetic field
crosses over from a linear temperature dependence $\Delta \rho \sim
T\sqrt{\rho_0}$ to a resistivity linear in $B$ and independent of $T$
and $\rho_0$. At higher magnetic fields $\Delta \rho$ saturates at a
value proportional to $T^{1.5}$ or $T^2/(p-p_c)$.  Deviations from
scaling, the interplay of orbital and spin contributions of the
magnetic field and experimental test of the spin-fluctuation model are
discussed in detail.
\end{abstract}
}

\section{Introduction}

One of the most important unresolved questions in the field of
strongly correlated metals is the stability of the Fermi liquid.  In
one dimension, the Fermi liquid is destroyed by strong quantum
fluctuations. The strange properties of the normal phase of
high-temperature superconductors \cite{andersonBook} as well as the
observation of an unexpected metal-insulator transition in Si-MOSFETs
\cite{Kravchenko} and other systems have cast doubts on the
applicability of Landau's Fermi-liquid paradigm in two dimensions.

Even in three dimensions ($3d$) a number of heavy-fermion metals near
an antiferromagnetic (AFM) quantum critical point (QCP) were shown to
display striking deviations from conventional Fermi-liquid behavior
\cite{loehneysen,review,julian,mathur,gegenwart,grosche,hauser,heuser,wilhelm,trovalli,maple}. The
main aim of this paper is to develop a detailed transport theory which
allows to decide experimentally whether the non-Fermi liquid behavior
near the QCP in weakly disordered metals can be explained by a nearly
AFM Fermi liquid.

It is believed that the magnetic transition in heavy Fermion systems
results from the competition between the Kondo screening of magnetic
moments and the AFM correlations induced by the RKKY interaction
\cite{doniach}. Pressure, magnetic fields or doping can
influence this interplay of screening and magnetism. This allows to
fine-tune systems directly to the QCP where the strongest non-Fermi
liquid (NFL) effects are found; a schematic phase diagram is shown in
Fig.~\ref{phasediaHot.fig}.  In particular, the resistivity rises
as a function of temperature with exponents smaller than $2$, the
specific heat coefficient diverges or shows a $\sqrt{T}$ cusp, and the
susceptibility shows anomalous corrections of the form $T^{\alpha}$,
with $\alpha < 1$.  We stress that both magnetic and non-magnetic
phases are heavy-Fermion metals which display, for example, the
characteristically large specific heat. The discussion of this paper is restricted to the behavior in the paramagnetic phase close to the QCP.

Despite the growing amount of experimental data, a common agreement on
 the origin of the observed non-Fermi liquid behavior is still
 lacking. At present it is unclear, whether a single mechanism is
 responsible for the observed behavior as different exponents have
 been reported in various compounds.  At least three different
 theoretical scenarios have been proposed and are discussed in
 literature.

The first scenario is based on the assumption that in a heavy
Fermion system below a scale $T_K$ the low energy excitations are
(heavy) quasi particles and their collective excitations.  In this
case the QCP should be in the same universality class as the
weak-coupling spin-density wave (SDW) transition in a Fermi liquid
studied by Hertz \cite{hertz,millis,moriya}. More precisely, the
so-called non-Fermi liquid behavior near the QCP is determined by the
mutual interaction between Landau damped spin fluctuations and
inelastically scattered quasi-particles in a nearly antiferromagnetic
Fermi liquid.

 Alternatively, one can envision a situation where the Kondo effects
breaks down directly at the transition, e.g. due to strong magnetic
fluctuations \cite{schroeder,si,coleman,continentino}. A Fermi liquid
description of the transition is then not possible and it has been
speculated that {\em local} fluctuations play the dominant role in
this scenario \cite{si,coleman}.

Independent of the precise nature of the QCP, sufficiently
strong disorder will certainly change the nature of the transition.
Indeed, if typical fluctuations of the effective Ne\`el temperature
due to disorder are large compared to the distance of the quantum
critical point (Harris criterion), disorder effects are
important. Perhaps more importantly, even at some distance form the
QCP, in the non-magnetic phase, rare configurations of impurities can
lead to a small magnetically ordered regions which can dominate some
of the thermodynamic properties in a finite region around the QCP
(Griffiths-McCoy singularities) \cite{griffiths,neto,belitz}.

Up to now, a comprehensive theoretical description exists only for
the weak-coupling spin-density wave transition \cite{hertz,millis}. As
this theory is above its upper critical dimension, essentially all
low-energy properties can be determined analytically. It was realized
only recently \cite{roschPRL,roschLT,hlubina} that the calculation of
transport quantities is quite subtle and earlier predictions
\cite{moriya,hlubina} are not valid in weakly disordered materials.
The complications arise because the scattering from spin-waves is
extremely anisotropic and effective only in small areas on the Fermi
surface.  Therefore the transport properties strongly depend both
qualitatively and quantitatively on how {\em other} scattering
mechanisms redistribute quasi particles and scatter them into these
small regions. 

The resistivity is often the most sensitive experimental probe to
study the QCP, and in a number of (pressure tuned) systems it is the
only available quantity. As the effects of strong disorder are poorly
understood at present, it is important to study very clean systems.
The goal of this paper is therefore to establish a set of predictions
for the resistivity within a SDW approach for weakly disordered metals
close to an AFM QCP. The interplay of the relevant scattering
mechanisms is studied within a semi-classical approach using a
Boltzmann equation.  The model and the Boltzmann equation is defined
in section \ref{modelS}. In section \ref{boltzsection} the analytic
solution of the Boltzmann equation is presented in the limit of low
temperature and weak, but finite disorder. The scaling properties
close to the quantum critical point are emphasized and numerical
solutions of the transport equations are used to analyze crossover
regimes and non-universal behavior.  Orbital effects of a magnetic
field are studied in section \ref{magentosection}.  Large non-linear
effects can be used as a tool to investigate the strongly anisotropic
scattering from the spin waves.  We will argue in section
\ref{discussionSection} that these calculations easily explain some of
the key observations in high purity single crystals close to an AFM
QCP: the temperature dependence of the resistivity changes from
$\Delta \rho \propto T^{1.5}$ to $\Delta \rho \propto T$ for cleaner
and cleaner systems \cite{julian,mathur,grosche}. We propose a number
of stringent tests for our picture and discuss how the orbital effects
of a magnetic field can be separated from spin-contributions.  We
conclude in section~\ref{conclusions} by commenting on the relevance
of this work to a wider class of problems.

\section{Resistivity at the QCP}
\subsection{Model}\label{modelS}
\begin{figure}[t]
  \centering
 \includegraphics[width=0.95 \linewidth]{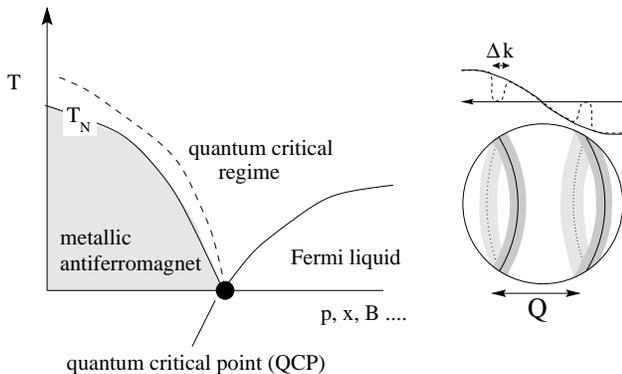}
\caption[]{Left figure: In a metallic AFM the magnetic order is
suppressed as a function of some control parameter which can e.g. be
pressure, doping or magnetic field. Deviations from the usual Fermi
liquid behavior show up close to the QCP. Right figure: Near the
transition to an antiferromagnet with ordering vector ${\vec{Q}}$, the
scattering on the Fermi-surface is enhanced along ``hot lines''
connected by ${\vec{Q}}$. This strong scattering equilibrates the
distribution function (shown for an electric field $\vec{E}$ parallel
to $\Q$) in a region of width $\Delta k$ (see the main
text and appendix \ref{appVar} for details). }
\label{phasediaHot.fig}
\end{figure}

Following Hertz \cite{hertz,millis} a spin-density
 antiferromagnetic
transition in a metal can be modeled by an effective Ginzburg-Landau
theory  defined by the action
\begin{eqnarray}\label{S}
S&=&\frac{1}{\beta} \sum_{\w_n,\vec{q}} \Phi_{\vec{q},i \w_n}^*
 \left(r+\frac{(\vec{q} \pm \vec{Q})^2}{q_0^2}+\frac{|\w_n|}{\Gamma}
 \right) \Phi_{\vec{q},i \w_n} \\ && + U \int_0^\beta d \tau \int d^3
 \vec{r} |\Phi(\vec{r},\tau)|^4 \nonumber \end{eqnarray} where $\Phi$
 is the order parameter which fluctuates in space and (imaginary)
 time, $\beta=1/k_B T$ is the inverse temperature, $\w_n= 2 \pi
 n/\beta$ are bosonic Matsubara frequencies. The fluctuations are
 strongest near the ordering vector $\vec{Q}$ of the AFM.  The mass
 $r$ controls the distance from the QCP.  The Landau damping term
 linear in $|\w|$ is due to the effective scattering of quasi-particles from
spin-fluctuations. A large number of  particle-hole pairs
is created by a spin-fluctuation if  $\Q$
 connects different parts of the Fermi surface ($Q < 2 \kf$, see
 Fig.~\ref{phasediaHot.fig}).  We will not discuss the singular case
 ''$\vec{Q}=2 \kf$'' \cite{millis,2kf} where the Fermi velocities
 $\vec{v}_\k$ and $\vec{v}_{\k\pm \Q}$ are parallel or ''$\vec{Q}>2
 \kf$'' where the spin-fluctuations do not couple directly to the
 quasi particles.

Due to the ohmic Landau damping, the effective field theory (\ref{S})
is characterized by a dynamic exponent $z=2$ ($\w \sim q^{z}$) and is
for $d=3$ above its upper critical dimension \cite{hertz,millis},
$d+z>4$. Therefore the dynamic susceptibility is mainly determined by
the gaussian part of (\ref{S}) and the spin-fluctuation spectrum
in the paramagnetic phase can be modeled by \cite{hertz,millis,moriya}
\begin{eqnarray} \label{chi}
\chi_{\vec{q}}(\w)=\chi_{-\vec{q}}(\w)\approx \frac{1}{1/(q_0
\xi)^2+\w_{\vec{q}}- i \w/\Gamma}, \end{eqnarray}
 where $q_0\approx
\kf$ and $\Gamma$ are characteristic momentum and energy scales and
$\xi$ is the correlation length of the spin fluctuations. The
dispersion $\w_\q \ge 0$ vanishes at the ordering wave vectors $\Q_i$
and varies on the scale $q_0$.  For
simplicity, we will mainly consider ``isotropic'' momentum dependences
\cite{2dim} $\w_{\q}\approx ((\vec{q}\pm \Q_i)/q_0)^2$ (appendix
\ref{appendixModelA} defines the models used for numerical
calculations more precisely).  Moderate anisotropy influences our
results only slightly, as will become clearer below.
$\xi$ is the AFM correlation length which
diverges \cite{hertz,millis}    at the QCP 
as $1/ \xi^2 \propto U T^{3/2}$.  For the
purposes of our numerical calculations we set 
\begin{equation} 1/(q_0
\xi)^2=r+c (T/\Gamma)^{3/2} \end{equation} with $c=1$. The
$T$-dependence of $\xi$ does not affect the low-temperature properties
(below Eqn. (\ref{magneto2}) a remarkable exception is discussed).  We
use $c=1$ to model the (non-universal) destruction of the spin
fluctuations at the temperature scale $\Gamma$. The parameters in
(\ref{chi}) can directly be obtained from inelastic neutron
scattering. Typically, it is expected that in a heavy Fermion system,
$\Gamma$ is of the order of the coherence-temperature or Kondo energy
$T_K$.  Close to the QCP, $r$ is a linear function of the
tuning parameter, e.g.  $r \propto p-p_c$ in a pressure tuned
experiment.

Transport can be treated within a simple (quantum-) Boltzmann
approach, because the spin-spin interactions are irrelevant in the
renormalization group sense and because furthermore the concept of
Fermi quasi particles is still valid \cite{hotLifetime} in 3
dimensions. For small and static electric fields the transport
equations take the usual form of fermions scattering from bosonic
excitations. These equations are further simplified by the assumption
that the spin-fluctuations stay in equilibrium, i.e. we neglect drag
effects. This approximation implicitly assumes the presence sufficient
momentum relaxation e.g. by strong Umklapp scattering.  In the linear
response regime, the quasi particle distribution
$f_{\vec{k}}=f^0_{\vec{k}}-\Phi_{\vec{k}} (\partial
f^0_{\vec{k}}/\partial \epsilon_{\vec{k}})$ is linearized around the
Fermi distribution $f^0_{\vec{k}}$ and the collision term reads
\cite{hlubina}
\begin{eqnarray} 
\left.\frac{ \partial f_{\vec{k}}}{\partial t}\right|_{\text{coll}} \!\!\!
&=&  
\sum_{\vec{k}'} \frac{f^0_{\vec{k}'} (1-f^0_{\vec{k}})}{T} 
(\Phi_{\vec{k}}-\Phi_{\vec{k}'})\nonumber \\
&& \hspace{-1.5cm} \times
\left[ 
g_{\text{imp}}^2 
\delta(\epsilon_{\vec{k}}-\epsilon_{\vec{k}'})+\frac{2 g_S^2}{\Gamma}
n^0_{\epsilon_{\vec{k}}\!-\!\epsilon_{\vec{k}`}}
\IM \chi_{\vec{k}-\vec{k}'}(\epsilon_{\vec{k}}\!-\!\epsilon_{\vec{k}`})
\right] \label{coll}
\end{eqnarray}
Here $g_{\text{imp}}^2$ and $g_S^2$ are transition rates for
impurity scattering and inelastic scattering from spin fluctuations,
respectively, and $n^0_{\w}$ is the Bose function.

In the following we will show that the interplay of the two scattering
mechanisms is highly non-trivial because of their completely different
momentum dependence. On the one hand, impurity scattering is very
isotropic and is equally efficient on the whole Fermi surface.
Therefore, the distribution function in the presence of a small
electric field $\vec{E}$ is smooth $\Phi_\k \propto \vec{v}_\k
\vec{E}$ if impurity scattering dominates (solid line in the right
part of Fig.~\ref{phasediaHot.fig}).  On the other hand the AFM
susceptibility is strongly peaked around the ordering vectors $\Q_i$.
Accordingly, a quasiparticle with an energy $\epsilon_{\k} \approx
\mu$ will scatter efficiently from the spin fluctuations only near
``hot lines'' on the Fermi surface where $\epsilon_{\k_H} \approx
\epsilon_{\k_H \pm \Q_i}\approx \mu$. This strong scattering tends to
equilibrate the distribution function at points $\k_H$ and $ \k_H \pm
\Q_i$ in a region of width $\Delta k$ (dashed line in
Fig.~\ref{phasediaHot.fig}), where $\Delta k$ depends on the relative
strength of impurity- and spin-fluctuation scattering. The temperature
dependence of $\Delta k$ and the distortion of the distribution
function close to the hot lines in a magnetic field is the main origin
of the anomalous transport properties which are discussed in this
paper (see appendix~\ref{appVar} for a simple qualitative calculation of
$\Delta k$).

For convenience, we will use in the following the dimensionless
quantities $t$, $x$, $r$ and $b$ to measure the effective temperature,
the amount of disorder, the distance from the quantum-critical point
and the strength of the magnetic field.
\begin{eqnarray}
t=\frac{T}{\Gamma}, \quad x=
\frac{\pi g_{\text{imp}}^2}{2 g_S^2}=\frac{\rho_0}{\rho_M}\approx 
\frac{1}{\text{RRR}}, \nonumber \\
 r=\frac{1}{(q_0 \xi(t=0))^2} \propto p-p_c, \quad b=\frac{B}{B_0}
 \label{defConst}
\end{eqnarray}
where $\rho_0=x \rho_M$ is the residual resistivity and $\rho_M$ is a
typical high-temperature ($t\approx 1$) resistivity which is
defined below.  One can approximately identify $x$ with the
inverse of the residual resistivity ratio (RRR).  $B_0$ is the typical
magnetic field, which is necessary to see Shubnikov-de Haas
oscillations at $t=1$
\begin{eqnarray} \label{Brho0}
\rho_M=\frac{3 \hbar g_S^2}{\pi e^2 v_F^2}, \quad 
B_0=\frac{2 g_S^2 q_0^d}{e v_F^2 (2 \pi)^d}
\end{eqnarray}
The Fermi velocity $v_F$ is defined by an average over the Fermi
surface $v_F^2/3=2/N_F \surfint d\vec{k} (\vec{v}_{\k}
\vec{n})^2/(v_{\vec{k}} (2 \pi)^3)$ where $\vec{n}$ is a unit vector
in the direction of the electric field. $\surfint d\vec{k}/v_\k=
\int \delta(\epsilon_\k-\mu) d^d \k$ is
an integral over the Fermi surface in $d$ dimensions and $N_F=2 \surfint
d\vec{k}/(v_{\vec{k}} (2 \pi)^3)$ the density of states.

\subsection{Boltzmann equation and scaling} 
\label{boltzsection}

The linearized Boltzmann equations with the collision term (\ref{coll})
can be written in the following form
\begin{eqnarray} \label{boltz}
\vec{v}_\k \vec{E}+ (\vec{v}_\k \times \vec{B}) \partial_{\k} \Phi_\k
= \surfint F_{\k \k'} (\Phi_\k-\Phi_{\k'}) \frac{d \k'}{v_{\k'} (2 \pi)^3}
\end{eqnarray}
where an integration over directions perpendicular to the Fermi
surface has already been performed and all $\k$-vectors and
integrations are restricted to the Fermi surface $\surfint d \k/v_\k
\equiv \int d^d \k \delta(\epsilon_\k-\mu)$ where $d$ is the number of
dimensions. Currents are calculated from $j_i = \surfint v_i
\Phi_{\k} d \vec{k}/(v_{\vec{k}} (2 \pi)^3)$.

$F_{\k \k'}$ includes contribution both from elastic scattering from
impurities and inelastic scattering from spin fluctuations with
\begin{eqnarray} 
F_{\vec{k}\vec{k}`}&=&g_{\text{imp}}^2+\frac{2 g_S^2}{\Gamma T} 
\int_0^{\infty}
\w n^0_{\w} [ n^0_{\w}+1] \IM \chi_{\vec{k}-\vec{k}`}(\w) d \w. \label{F}
\\
&=&  g_{\text{imp}}^2+2 g_S^2 I(r/t+\w_q/(q_0^2 t)). \nonumber
\end{eqnarray}
$I(y)$ is defined by equation (\ref{F}) and (\ref{chi}) and is
 asymptotically given by $I(y \to 0)\approx \pi/(2 y)$ and $I(y \to
 \infty )\approx \pi^2/(3 y^2)$.

For large temperatures and extremely large magnetic fields the
solution of (\ref{boltz}) depends on all details of the Fermi surface
and only numerical solutions are possible. However, close to the
quantum critical point the resistivity shows scaling behavior and
depends only slightly on the geometry and the Fermi velocities along
the hot lines.  In the following we will derive an approximate
analytic solution of the Boltzmann equation in this regime, the
approximations used are exact in a well-defined limit of weak
disorder, low temperature, weak coupling and small magnetic fields
which will be specified below.  Some of our qualitative results,
especially in the absence of a magnetic field, can be derived by using
a simple variational approach (appendix~\ref{appVar}) or even in a
simple relaxation time approximation which we present in appendix
\ref{relaxTime}. The relaxation time approximation or, equivalently,
the omission of vertex correction is completely wrong in certain
geometries and leads to large errors especially for the Hall
coefficient as explained in appendix \ref{relaxTime}.

At $T=0$ the scattering is purely elastic and the quasiparticle
distribution function is given by $\Phi^0_\k(\vec{B}) \equiv
\Phi_\k(B,T=0)$ with $\Phi^0_\k(0)=e \vec{E}
\vec{v}_\k/(g_{\text{imp}}^2 N_F)$.  Our strategy is to keep track of
the change of $\Phi_\k$ close to the hot lines, where the strong
spin-fluctuations tend to equilibrate the distribution function at the
points $\k$ and $\k\pm \Q_i$ (Fig.~\ref{phasediaHot.fig}). For
sufficiently small temperatures $t^2 < x$ the distribution function
will remain undisturbed in the ``cold regions'' far away from the hot
lines where the inelastic scattering is of order $t^2$ and the elastic
scattering dominates.  It is therefore convenient to write the
distribution function in the following form \cite{defProblem}
\begin{eqnarray}\label{pkdef}
\Phi_\k \approx (1-p_\k) \Phi_{\k}^0+ p_\k
\Phi_{\k\pm \Q_i}^0.
\end{eqnarray}

For strong spin-scattering $p_\k$ approaches a value close to $1/2$
 near the hot lines but vanishes further away. It is helpful to use a
 coordinate system, where vectors on the Fermi surface  $\k=\k_H+\k_{\perp}$
are split up in
 a vector $\k_H$ on the hot line and a perpendicular vector
 $\k_{\perp} \perp \k_H,\vec{v}_\k$.  After a
 rescaling of the momenta using $\k=\kappaV q_0 \sqrt{t} $,
in the limit $t<\sqrt{x}$ the
 Boltzmann equation takes the following form
\begin{eqnarray}
t^{(d-1)/2} \int& d\kappaV` & (p_\kappaV+p_{\k_H\pm \Q+\kappaV'}-1) 
I(r/t+M_{\kappaV \kappaV'}) \nonumber\\
&& + \sqrt{\alpha_H} c_{\k_H} x p_\kappaV = c'_{\k_H}
 \frac{b}{\sqrt{t}} (\hat{\vec{v}}_\k 
\times \hat{\vec{B}}) \partial_{\kappaV} 
p_\kappaV  \label{boltzScaling}
\end{eqnarray}
with $I(y)=\int_0^{\infty} dz z^2 n^0(z)(1+n^0(z))/(y^2+z^2)$,
$\kappaV'=\kappaV'_{\|}+\kappaV'_{\perp}$, $M_{\kappaV
\kappaV'}={\kappa'_{\|}}^2+ (\kappaV'_{\perp}-\kappaV_{\perp})^2$,
$c_{\k_H}= (2 \pi)^d N_F \sqrt{v_{\k_H}v_{\k_H\pm \Q}} /(\pi q_0^2)$,
$c'_{\k_H}= v_{\k_H}v_{\k_H\pm \Q} /v_F^2$ and \cite{alphaH1}
$\alpha_H=v_{\k+\Q}/v_{\k}$. $\hat{\vec{B}}$ and $\hat{\vec{v}}_\k$
are unit-vectors in the direction of $\vec{B}$ and $\vec{v}_\k$,
respectively. The strength of the magnetic field is measured in units
of a typical field $B_0$ defined in (\ref{defConst}) and
(\ref{Brho0}).  One can think of $b$ as the typical angle by which an
electron is deflected within a typical scattering time $\tau_M=(2
\pi)^d v_F/(2 g_s^2 q_0^{d-1})$.  We are only considering magnetic
fields with $b \ll x<1$ (i.e. $\w_c \tau \ll 1$), where
Shubnikov-de-Haas oscillations are absent at low temperatures.  In
(\ref{boltzScaling}) we have neglected sub-leading contributions
e.g. of the form $p_{\k} \partial_\k \Phi_{k}^0$, which are small in
this regime because $p_{\k}$ varies stronger with momentum than
$\Phi_{k}^0$ and all inelastic corrections come from small areas on
the Fermi surface. This is not true in the ultra-clean limit $x \ll
t^2$ which has to be covered separately (e.g. in section
\ref{hallSection}).  For completeness we discuss some of our result in
$d$ dimensions, however, we are mainly interested in three
dimensions. Note that the Boltzmann equation approach presumably
breaks down in $d=2$ as the ``hot'' electrons acquire a lifetime
\cite{chubukov} $\propto \sqrt{T}$, but it was nevertheless widely
used in literature \cite{hlubina,pines}.  Qualitative features like
the interplay of hot and cold regions might still be correctly
described by our approach even in $d=2$.

From (\ref{boltzScaling}) it is clear that only the combinations 
$t^{(d-1)/2}/x$, $t/r$ and $b/(x \sqrt{t})$ 
determine $p_{\kappaV}$. Accordingly, the resistivity
obeys a scaling relation 
\begin{eqnarray}\label{scaling1}
\frac{\Delta \rho_{i j}}{\rho_M}&\approx& t^{\frac{d}{2}} 
f_{i  j}\!\left(\frac{t^{
\frac{d-1}{2}}}{x},
\frac{r^{\frac{d-1}{2}}}{x},\frac{b}{x \sqrt{t}}\right) 
\end{eqnarray}
for $t \ll \sqrt{x} \ll 1$, $r\ll1$ and
sufficiently small $b$,
where $f$ depends smoothly on the details of the Fermi surface near
the hot spots and the directions of magnetic and electric fields. 
 $\rho_M$ is defined in Eqn. (\ref{Brho0}).
The relation (\ref{scaling1}) becomes exact in the scaling limit
\begin{eqnarray}\label{scalinglimit}
t,x,b,r \to 0,\quad  \frac{t^{(d-1)/2}}{x}, 
\frac{r}{t}, \frac{b}{x \sqrt{t}}
 \to const.
\end{eqnarray}
The scaling limit serves as a convenient regime for
analytic calculations. An experimental verification of the predicted
scaling
would be the most precise test of the underlying models. We will
calculate the scaling function $f$ in the following paragraphs in
detail.  However, for some applications, e.g. the Hall effect
discussed in section \ref{hallSection}, the deviations from scaling
are important and are calculated from a full numerical solution of the
Boltzmann equation (\ref{boltz}).  A
reader  interested only in the qualitative results can jump
directly to section \ref{overview}, where a short overview of the
asymptotic behavior of $f_{i i}$ is given (the off-diagonal components
$f_{i \neq j}$ vanish in the scaling limit, see section
\ref{hallSection} where the Hall effect is discussed).

The two dimensional coupled integro-differential equations
(\ref{boltzScaling}) can be considerably simplified by realizing, that
$p_\k$ varies smoothly parallel to the hot lines but the kernel
$I(\kappaV'_\|)$ restricts the integrations to small values of
$\kappaV'_\|$. This allows us to replace $p_{\k_H \pm \Q+ \kappaV'}
\equiv \tilde{p}_{\kappa_\|+\kappa'_\|, \kappa'_\perp }$ by
$\tilde{p}_{\kappa_\|, \kappa'_\perp }$ in (\ref{boltzScaling}) and to
perform the $\kappa'_\|$ integration. For the same reason we can
neglect the contribution proportional to $\partial_{\kappa_{\|}}
p_{\kappaV}$. All these approximations are valid in the scaling limit
(\ref{scalinglimit}).  It is therefore sufficient to solve a family of
two coupled one-dimensional integral equations which depend
parametrically on $\k_H$:
\begin{eqnarray}\label{boltz1d}
t^{(d-1)/2} \int& d\kappa' & (p_\kappa+\tilde{p}_{\kappa'}-1) 
G_{\kappa \kappa'} \nonumber\\
&& + \sqrt{\alpha_H} c_{\k_H} x p_\kappa = 
c'_{\k_H} \cos \theta_B \frac{b}{\sqrt{t}} 
 \partial_{\kappa} p_\k \\
t^{(d-1)/2} \int& d\kappa' & (\tilde{p}_\kappa+p_{\kappa'}-1) 
G_{\kappa \kappa'} \nonumber\\
&& + \sqrt{\frac{1}{\alpha_H}} c_{\k_H} x \tilde{p}_\kappa =
 c'_{\k_H} \cos \theta_B \frac{b}{\sqrt{t}} 
 \partial_{\kappa} \tilde{p}_\k \nonumber
\end{eqnarray}
where $\theta_B$ is the angle between $\vec{B}$ and a vector parallel
to the hot lines at $\k_H$ and 
\begin{eqnarray}
G_{\kappa_\perp \kappa'_{\perp}}&=&\int
d^{d-2} \vec{k}_{\|} I(r/t+M_{\kappaV \kappaV'})\\
&\approx& \left\{ \begin{array}{ll}\displaystyle
\frac{1}{2 \sqrt{a_{\kappa \kappa'}}} & a_{\kappa \kappa'}\ll 1\\[1em]
\displaystyle 
\frac{\pi}{6 a_{\kappa \kappa'}^{3/2}} & a_{\kappa \kappa'} \gg 1
\end{array}\right. \text{for } d=3.\label{Gasypt}
\end{eqnarray} 
with $a_{\kappa \kappa'}=r/t+\kappa^2+\kappa'^2-2 \kappa \kappa' \cos
\phi$, where $\phi \neq 0$ is the angle between $\vec{v}_{\k_H}$ and
$\vec{v}_{\k_H\pm \Q}$.  The equation for $\tilde{p}_{\kappa}$ is
obtained by replacing $\alpha_H$ by $1/\alpha_H$ and exchanging
$p_{\kappa}$ and $\tilde{p}_{\kappa}$ in (\ref{boltz1d}). The boundary
condition for (\ref{boltz1d}) are
$p_{\kappa},\tilde{p}_\kappa \to 0$ for $\kappa \to \pm \infty$
 by construction.

Currents are calculated from 
\begin{eqnarray} \label{current}
j^i&=&j^i_0+\\
&& 2 \sum_i \oint_i \frac{d \vec{k}_H}{v_{\k_H} (2 \pi)^3} v^i_{\k_H} 
(\Phi_{\k_H}^0-\Phi_{\k_H+\Q}^0)q_0 \sqrt{t} \int d \kappa
p_{\kappa}\nonumber
\end{eqnarray}
where $\vec{j}_0=2 \surfint \vec{v}_{\k} \Phi_{\k_H}^0 d^3\k/(v_\k (2
 \pi)^3)$ is the current in the absence of inelastic scattering,
 $\oint_i d \vec{k}_\|$ denotes the line integral along the $i$th hot
 line and the expression is summed over all hot lines $i$. 
Note that $p_\kappa$ is a function of $\k_H$.  In the scaling
 limit $b \ll x$, the magnetic field dependence of
 $\Phi_{\k_H}^0\propto \vec{v_{\k_H}} \vec{E}$ can be neglected in the
 second term in (\ref{current}). Using the symmetries of the Boltzmann
 equations one can derive the convenient expression
\begin{eqnarray}\label{rhoij}
\frac{\Delta \rho_{i j}}{\rho_M} \approx x \frac{
\sum_i \oint_i 
\frac{d \vec{k}_\|}{v_{\k_H}} \Delta v^i \Delta v^j
(q_0 \sqrt{t}) \int  d \kappa p_\kappa}{
\surfint (v_\k^i)^2 \frac{d \vec{k}_\|}{v_{\k}}}.
\end{eqnarray}
with $\Delta v^i=v^i_{\k_H}-v^i_{\k_H \pm \Q}$.

\subsection{Transport for $B=0$}
In the absence of a magnetic field, two different regimes emerge
depending on whether the elastic scattering dominates close to the hot
line \cite{roschPRL,roschLT}. If impurity scattering dominates 
($t < \min[x,\sqrt{r x}]$), it
smears out the quasi particle distribution ($\Delta k \approx 0$ in
Fig.~\ref{phasediaHot.fig}).  $p_{\kappa}$ in (\ref{boltz1d})
is small and in leading order in $p_{\kappa}$ one obtains
\begin{eqnarray} \label{rhoImp}
\frac{\Delta \rho}{\rho_M}&\approx&
t^{3/2} h\!\left(\frac{t}{r}\right)  \frac{3 q_0^3}{8 \pi^4 (v_F N_F)^2 }
\nonumber \\ 
&& \times \sum_i 
\oint_{i} d \k_{\|} \frac{((\vec{v}_{\k_{\|}}-\vec{v}_{\k_{\|} \pm \Q_i}) 
\vec{n})^2}
{|\vec{v}_{\k_{\|}} \times \vec{v}_{\k_{\|}\pm \Q_i}|} 
 \\
h\left(\frac{t}{r}\right) &=& \frac{\pi}{2} \int_0^{\infty}
 y n^0(y)(1+n^0(y))  \IM \sqrt{\frac{r}{t}+ i y} \nonumber \\
&\approx& \left\{ \begin{array}{ll}
\frac{\pi \zeta(3/2) \Gamma(5/2)}{2 \sqrt{2}} 
 &, t >r  \\[.3cm]
\frac{\pi^3}{12}\sqrt{t/r}
&, t< r  
\end{array} \right.
\end{eqnarray}
In this dirty limit, we recover the well-known $T^{1.5}$ resistivity at
the QCP. This result is usually derived by an average over the
inelastic  scattering rate\cite{moriya}. We want to stress that this
approximation can be applied {\em only} in the presence of strong
impurity scattering and gives wrong results for cleaner systems.

\begin{figure}[t]
  \centering
 \includegraphics[width=0.75 \linewidth]{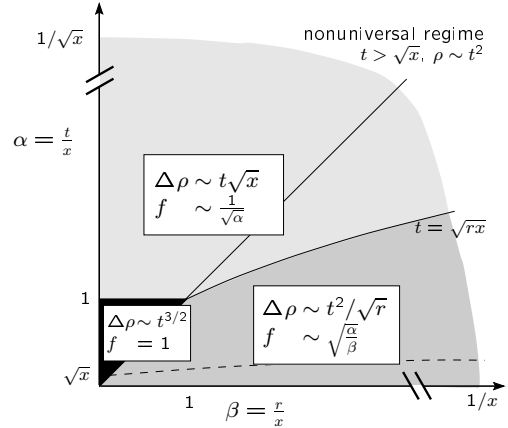}
\caption[]{In the scaling limit $t,x,r \to 0$, $t/x,r/x \to const.$
the resistivity is ``universal'' $\Delta \rho/\rho_M = t^{3/2}
f(t/x,r/x)$ for $t<\sqrt{x}, r<1$, where $t\propto T$, $x\propto
\rho_0$ and $r\propto p-p_c$ measure the temperature, the amount of
disorder and the distance from the QCP. The plot shows the qualitative
behavior of the scaling function $f[\alpha,\beta]$ in the various
regimes. For $\min[x, \sqrt{r x}<t<\sqrt{x}]$ the resistivity rises
linearly with temperature. Thermodynamic quantities show a crossover
to Fermi liquid behavior at the scale $t=r$ (straight line), while in
transport a $T^2$ behavior is recovered at a much smaller scale
$\min[x, \sqrt{r x}]$. The dashed line serves as a reminder that at
lowest temperatures effects which are not included in our approach
become important, e.g.  interference effects of disorder and
interactions \cite{altshuler} or a disorder induced change of the spin
fluctuation spectrum \cite{belitz}.  }
\label{scaleRho}
\end{figure}
To derive (\ref{rhoImp}) we considered for simplicity
 isotropic spin fluctuations. In
the case of a moderate anisotropy our results have to be changed
slightly. For example, if the anisotropy of the spin fluctuations is
described by $\vec{q}_i g_{ij}\vec{q}_j/q_0^2$, then the denominator
of the integral in (\ref{rhoImp}) has to be replaced by 
$v_{\k_H} v_{\k_H \pm \Q}\det^{1/2}
\sum_{ij} \hat{\k}_i^{\alpha} g_{ij} \hat{\k}_j^{\beta}$ where the
$\k^1$ is a unit vector parallel to the hot line  and 
$\k^2, \k^3$ are  unit vectors perpendicular to $\k^1$ and either
$\vec{v}_{\k_H}$ or $\vec{v}_{\k_H\pm \Q}$. The determinant is always
finite because we do not consider the case
 $\vec{v}_{\k_\|} \| \vec{v}_{\k_\|+\Q}$ (``$\Q= 2 \kf$'').

\begin{figure}[t]
  \centering
 \epsfig{width=0.7 \linewidth,file=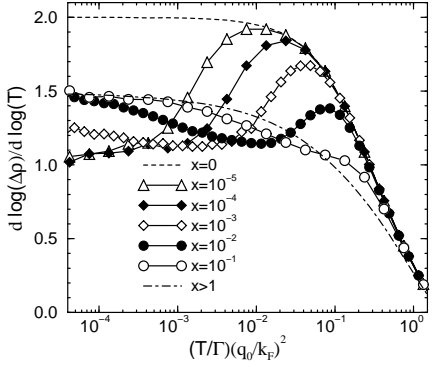}
\caption[]{Effective exponent of the resistivity at the QCP ($r=0$),
defined as the logarithmic derivative of $\Delta \rho(T)$. At very low
temperatures, the ``dirty-limit'' exponent $3/2$ is
recovered. However, in the experimentally accessible low temperature
regime smaller exponents are to be expected for rather clean system
($x<0.1$).}
\label{figExponent}
\end{figure}
In the limit $\sqrt{x}>t>\min[x,\sqrt{r x}]$,  the resistivity rises
{\em linearly} with temperature. The origin of this effect is that
the spin fluctuations equilibrate a region of
 width $\Delta k \approx t/\sqrt{x}$ 
(Fig.~\ref{phasediaHot.fig}) as we will show in Eqn. (\ref{boltztGTRx})
(or appendix~\ref{appVar}). Therefore
we obtain $\rho \sim x(1+t/\sqrt{x})$ or more precisely
\begin{eqnarray} \label{rholin}
\frac{\Delta \rho}{\rho_M}\approx&& t\sqrt{x} \biggl[
\frac{q_0^2  \sqrt{\pi/3}}{2 v_F^2 (2 \pi)^3 N_F^{3/2}}  \sum_{i} \oint_{i} 
d \k_{\|} \biggl[ \nonumber \\
&&  \frac{((\vec{v}_{\k_{\|}}-\vec{v}_{\k_{\|} \pm \Q_i}) 
\vec{n})^2}
{|\vec{v}_{\k_{\|}} \times \vec{v}_{\k_{\|}\pm \Q_i}|} 
\frac{\vec{v}_{\k_{\|} \pm \Q_i}^{5/4}}{\vec{v}_{\k_{\|}}^{3/4}} 
S_{\vec{v}_{\k_{\|} \pm \Q_i}/\vec{v}_{\k_{\|}}}^
{|\hat{\vec{v}}_{\k_{\|}\pm \Q_i} 
\hat{\vec{v}}_{\k_{\|}} |}\biggr]\biggr] \nonumber\end{eqnarray}
where the only slightly varying function $S_{\alpha}^a=\int_{-\infty}^{\infty}
 d \kappa p_\kappa(a,\alpha) \approx \pi$ is calculated from
the solution of the two coupled one-dimensional integral equations (see Eqn.
(\ref{boltztGTRx}))
\begin{eqnarray}
p_q+\int d q'  \frac{(p_q+
\tilde{p}_{q'}-1)(1-a^2)}{(q^2/\alpha+{q'}^2 \alpha-
2 q q' a)^{3/2}}&=&0\nonumber \\
\tilde{p}_q+\int d q'  \frac{(\tilde{p}_q+
p_{q'}-1)(1-a^2)}{(q^2 \alpha+{q'}^2/\alpha-
2 q q' a)^{3/2}}&=&0
\end{eqnarray}
which in the limit $a\to \pm 1$ are solved by 
$p_q=1/((1+\alpha)+q^2/2 \sqrt{\alpha})$ (see Fig.~\ref{distQClean}).

\begin{figure}[t]
  \centering
 \includegraphics[width=0.75  \linewidth]{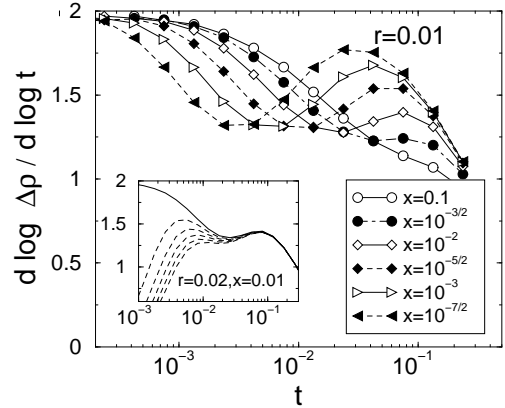}
\caption[]{Effective exponent, defined as the logarithmic derivative
of $\Delta \rho(T)$, for a fixed distance from the QCP $r=0.01$ and
various values of disorder. In high purity samples the crossover to an
exponent $2$ can be seen only at lowest temperature $t \ll
\min[x,\sqrt{r x},r]$ (\ref{rhoImp}). The bump at higher temperatures
is a precursor of the effect that in ultra clean samples, $x \ll t^2 \ll
1$, we expect $\Delta \rho \sim t^2$ (\ref{ultrat2}).  In the scaling
limit (\ref{scalinglimit}) the resistivity is linear in the
intermediate regime $\sqrt{r x}<t<\sqrt{x}$ (\ref{rholin}). For the
chosen parameters this exponent $1$ cannot be identified, however, a
pronounced regime with an effective exponent less than $1.5$ is
seen. The inset illustrates the problem to extract the exponent
\cite{mathur,grosche} from
the logarithmic derivative of $\Delta \rho=\rho-\rho_0$ if $\rho_0$
is not known. For $=0.02$ and $x=0.01$ the logarithmic derivative of
$\rho-(1-\epsilon) \rho_0$ for $\epsilon=2-10\%$ (dashed lines) is
compared to the ``true'' exponent for $\epsilon=0$ (solid line). }
\label{r01fig}
\end{figure}

\begin{figure}[t]
  \centering
 \includegraphics[width=0.75  \linewidth]{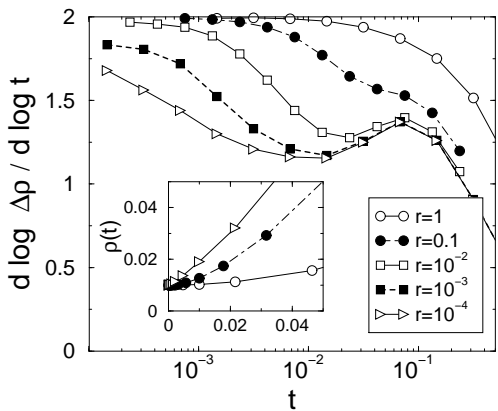}
\caption[]{Effective exponent, defined as the logarithmic derivative
of $\Delta \rho(T)$, for a relatively clean sample with $x=0.01$
and various values of $r \propto p-p_c$. The resistivity is
calculated from a numerical solution of the
Boltzmann equation for model A defined in appendix
\ref{appendixModelA}. The inset shows $\rho(t)$
for $r=0, 0.1, 1$.}
\label{x01fig}
\end{figure}

Omitting all prefactors we obtain in the scaling regime $r\ll 1, t <\sqrt{x}$
\begin{eqnarray}
\frac{\Delta \rho}{\rho_M}&\approx& t^{\frac{d}{2}} f\!\left(\frac{t^{
\frac{d-1}{2}}}{x},
\frac{r^{\frac{d-1}{2}}}{x}\right) \label{scaleB0} \\
&\sim& \left\{ 
\begin{array}{ll}
t^{d/2}& , r<t<x^{2/d-1} \\
t^{\frac{2}{5-d}} x^{\frac{4-d}{5-d}} &, 
\text{max}[x^{\frac{2}{d-1}},\sqrt{x} r^{\frac{5-d}{4}}]<t<\sqrt{x} \\
t^2/r^{2-\frac{d}{2}} &, t<\text{min}[r,\sqrt{x} r^{\frac{5-d}{4}}]
\end{array}
\right. \nonumber 
\end{eqnarray}
where for completeness we have given the result in arbitrary dimensions $d$
(see appendix \ref{appVar} for a simple derivation of these results).
A summary of the scaling behavior in $d=3$
is shown in Fig.~\ref{scaleRho}.
It is worthwhile to point out, that even away from the QCP, for
$r>0$, there is a large regime with a NFL resistivity (Fig.~\ref{r01fig}
and \ref{x01fig}). To actually observe the $T^2$-term one has to consider
very low temperatures $t<\sqrt{r x}$. This has to be compared to
thermodynamic quantities, where the crossover to Fermi-liquid behavior
is expected at a higher scale $t \sim r$. In a very
clean system these scales can be quite different and anomalous transport
should be observed in regimes where thermodynamic quantities 
display typical Fermi liquid properties.

For higher temperatures $t > \sqrt{x}$, larger resistivities $\Delta
\rho \gg \rho_0$ or further away from the QCP $r \sim 1$ the
resistivity is much less universal and one has to rely on numerical
solutions of the Boltzmann equation.  In the ultra-clean limit
$x<t^2\ll 1$, the distribution function $\Phi_\k$ is suppressed in a
large region around the hot lines \cite{hlubina} ($\Delta k \approx
O(k_F)$ in Fig.~\ref{phasediaHot.fig}). Therefore only non-singular
scattering dominates the resistivity and we obtain
\begin{eqnarray}\label{ultrat2}
\frac{\rho}{\rho_M} \approx t^2
\end{eqnarray} 
with a prefactor which to leading order is independent of the distance
from the QCP $r$ but depends on all details of the Fermi surface and
of the interactions {\em far} from the hot lines. Hlubina and Rice
\cite{hlubina} were the first to discuss this regime where
quantum-critical effects cannot be seen in the transport because the
long-living quasi particles in the ``cold regions'' short-circuit all
contributions from the hot lines.  Our numerical solutions, however,
suggest that this regime is experimentally not accessible.
Fig.~\ref{figExponent} shows the logarithmic derivative of $\Delta
\rho$, which defines an effective exponent $\alpha$, $\Delta \rho
\propto T^\alpha$.  Even for an extremely clean system with a residual
resistivity ratio (RRR $\approx 1/x$) of the order of $10^4$, there is
never an extended range of temperatures, where this effective exponent
is close to $2$. On the other hand we see that in moderately clean
systems with $x < 0.1$ there is always a regime, where the effective
exponent is smaller than $1.5$. Sufficiently clean systems clearly
show the linear resistivity predicted by (\ref{rholin}) over a large
range.

The full scaling function for $r=0$
and the deviations from scaling are shown in
Fig.~\ref{figScaling}.


\section{Magnetotransport}\label{magentosection}
The characteristic feature of the spin-density wave scenario is,
that only the quasi particles close to  ``hot lines'' are strongly affected
by the spin-fluctuations. This has to be contrasted with a situation
where the Kondo effect is destroyed close to the QCP which would affect
the full Fermi surface. It is therefore important to have a probe for testing
 the presence of sharp structures on the Fermi surface. 

The (nonlinear) magnetoresistivity can be used to investigate
precisely this question. By a magnetic field, the quasi particles are
driven parallel to the Fermi surface by the force $\vec{v_{\k}} \times
\vec{B}$.  Generally, non-linear effects in the magnetotransport show
up if the quasi particles are able to circle the Fermi surface without
being scattered, i.e. for $\w_c \tau \gtrsim 1$, where $\w_c$ is the
cyclotron frequency. However, if there is a small region of width
$\Delta k$ on the Fermi surface where the scattering is relatively
strong (or weak) then non-linear effects will become important for
$\w_c \tau \gtrsim \Delta k/k_F$.  Close the AFM quantum critical
point the width of hot lines $\Delta k$ vanishes in the
low-temperature limit and the quasi particles can be driven by a
moderate magnetic field over the hot lines. This will lead to large
effects in the magnetotransport. The magnetotransport in nearly
antiferromagnetic {\em two dimensional} metals has been studied before
\cite{pines,vertexHall}
in detail in the context of high-temperature
superconductors.  However, the authors did not
focus on the role of weak disorder, which we find at least in three
dimensions to be essential to describe the low-temperature the
magnetotransport both quantitatively and qualitatively.

\begin{figure}[t]
  \centering
 \epsfig{width=.8 \linewidth,file=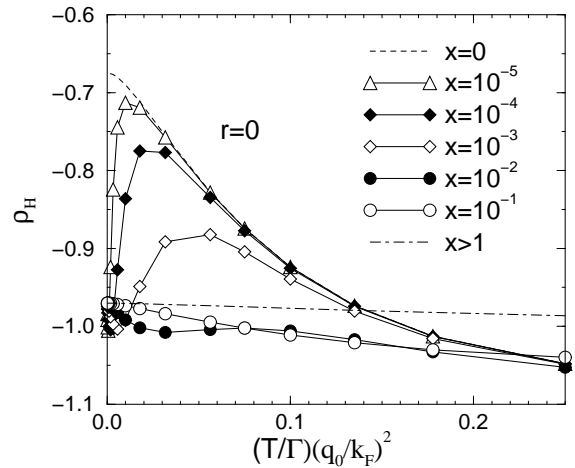}
\caption[]{Hall effect for model A, calculated from a numerical solution
of \ref{boltz}. Extremely small amounts of disorder strongly influence
the Hall effect. Very clean samples are necessary to observe this effect.
Both the sign and the size of the effect
 depend on details of the band structure.}
\label{figHall}
\end{figure}
 
\subsection{Hall effect}\label{hallSection}

The Hall effect is extremely sensitive to small amounts of disorder 
($x < t^2$) but essentially constant for $t \ll \sqrt{x}$.

This can be seen most easily from the symmetry properties of the
Boltzmann equation  (\ref{boltz1d}) in the scaling limit.
The left-hand side of (\ref{boltz1d}) is even in $\kappa$, while
the right hand side is odd and linear in the magnetic field. Therefore
$\int_{-\infty}^{\infty} p_\kappa$ is {\em even} in $\vec{B}$ and
there is no contribution to the Hall effect  from the hot
lines in the scaling limit (\ref{scalinglimit}) where (\ref{boltz1d})
and (\ref{rhoij}) are valid. This does not imply that the Hall effect
is unaffected by the singular scattering close to the QCP. It is on
the contrary an extremely sensitive probe to very small amounts of
impurity scattering.  The Hall constant at $T=0$ is different in the
ultra-clean and the dirty limit $R_H(t\to 0, x \ll t^2) \neq
R_H(t\to 0, x=const.)$. In the former limit the 
elastic impurity scattering dominates in the latter the
magnetic scattering. The  quasiparticle distribution 
$\Phi_{\vec{k}}$
has a completely different momentum dependence in these two limits.
One consequence is that  a different average
over the effective masses enter the Hall constant. Small amounts of
disorder $x \sim t^2$ change the distribution $\Phi_{\vec{k}}$ 
completely and give
rise to the strong features in the Hall effect shown in
Fig.~\ref{figHall}. The calculation is based on a numerical solution
of the full Boltzmann equations (\ref{boltz}) for small magnetic
fields. Even disorder at the level of e.g. $x=10^{-4}$ leads to a
strong structure in the Hall effect at quite high
temperatures. Extremely clean samples are necessary to observe
the strong temperature dependence of $R_H$.  Both the sign and the
size of these effects are non-universal and depend on the details of the
band structure.

\subsection{Magnetoresistivity -- small fields}
While all corrections to the Hall effect cancel in the scaling limit
(\ref{scalinglimit}), the magnetoresistivity is strongly influenced by
the hot lines and can be used to investigate the quantum-critical
properties in more detail.

We first discuss the dirty limit $x>t$ ($d=3$).  $p_\kappa$ is small
in this case allowing an  expansion of (\ref{boltz1d}) in
$p_{\kappa}$. In order $b^2$, the distribution function changes by
$\Delta p_{\kappa} \approx (b \cos \Theta_B c'_{\k_H}/(x \sqrt{t}
c_{\k_H} \sqrt{\alpha_H}))^2 \partial^2_{\kappa} p_{\kappa}^0$ where
$p_{\kappa}^0$ is the distribution function for $b=0$.  The
resistivity (\ref{rhoij}) is proportional to $\int p_{\kappa}$, but
$\int \Delta p_{\kappa}$  vanishes obviously. Therefore it is useful to
integrate (\ref{boltz1d})
\begin{eqnarray} \label{pint}
\int_{-\infty}^{\infty} \!\!\! d \kappa p_\kappa = \frac{t}{x 
\sqrt{\alpha_H} c_{\k_H}}
\int \!\!\! \int_{-\infty}^{\infty} \!\!\! d  \kappa d \kappa'
 G_{\kappa \kappa'} 
(1-p_{\kappa}-\tilde{p}_{\kappa'}).\nonumber \\
\end{eqnarray}
  We calculate the leading correction to $\int p_\kappa$
 from the right side of
(\ref{pint}), using that $p_{\kappa}^0 \approx t/(x \sqrt{\alpha_H}
c_{\k_H}) \int G_{\kappa \kappa'} d \kappa'$:
\begin{eqnarray}\label{pdiffK}
\Delta \int p_{\kappa} \approx \frac{1+\alpha_H^3}{\alpha_H}
\left(\frac{b \cos \Theta_B c'_{\k_H}}{x \sqrt{t} 
c_{\k_H}} \right)^2 \int d \kappa (\partial_\kappa  p_\kappa^0)^2
\end{eqnarray}
and obtain for $r<t<\min[x,\sqrt{r x}], b \ll x \sqrt{r}$
\begin{eqnarray} 
\frac{\Delta \rho}{\rho_M}&\approx& b^2 \frac{t^4}{x^3 r^{5/2}}  \biggl[
\frac{ \pi^7 q_0^9/(12 v_F^4)}{(N_F (2 \pi)^3)^5} \sum_{i} \oint_{i} 
d \k_{\|} \biggl[ \frac{v_{\k_{\|}}^3 \! + 
v_{\k_{\|}\pm \Q_i}^3}{v_{\k_{\|}}^4 v_{\k_{\|}\pm \Q_i}^4}\nonumber \\
&& \frac{((\vec{v}_{\k_{\|}}\!-\vec{v}_{\k_{\|} \pm \Q_i}) 
\vec{n})^2  ((\vec{v}_{\k_{\|}} \times \vec{v}_{\k_{\|}\pm \Q_i}) 
\hat{\vec{B}})^2}
{|\vec{v}_{\k_{\|}} \! \times \vec{v}_{\k_{\|}\pm \Q_i}|} 
\biggr]\biggr]  \label{magneto1}
\end{eqnarray}
where $\hat{\vec{B}}$ is a unit vector in the direction of the
magnetic field.  It is surprising that the leading low temperature
correction in this Fermi liquid regime $t \ll r$ is not proportional
to $T^2$ but starts with $T^4$. This is due to the above mentioned
cancellations and is actually valid only for $t>r \sqrt{x}$ as shown
below.

There are a number of other contributions to the
magnetoresistivity which vanish in the scaling limit but might be
dominating in the experimentally relevant regime.  One
important correction is due to the suppression of the AFM in a
magnetic fields, we will discuss these effects in detail in section
\ref{sectionSpinOrbit}.
Before analyzing the magnetoresistivity it is usefull to
subtract the $B$-dependence of the resual resistivity 
$\Delta \rho_0 \sim b^2/x$ which is the largest
correction to our results. Therefore we use in this section the definition
\begin{eqnarray}\label{delrho}
\Delta \rho=(\rho(b,t)-\rho(0,t))-(\rho(b,0)-\rho(0,0))
\end{eqnarray}
Other sub-leading contributions become important for higher magnetic
fields or in the case of Eqn. (\ref{magneto1}) for low temperatures.
They are missing in (\ref{boltzScaling}) because we neglected terms of
the form $\partial_{\k} (\Phi^0_{\k_H}-\Phi^0_{\k_H\pm \Q})$. We
estimate these corrections in the regime $\omega_c \tau \ll 1$ and
$\Delta \rho < \rho_0$ using Kohler's rule ($
\rho(T,B)-\rho(0,B)\propto B^2 \Delta \rho(T,0)/\rho(0,0)^2$).
In the regime discussed in  (\ref{magneto1}), they are of the order
of $\Delta \rho \propto b^2 t^2/(\sqrt{r} x^2)$
and  therefore (\ref{magneto1}) is valid only for
 $t > t^*= r \sqrt{x}$. In
the scaling limit (\ref{scalinglimit}), $t^*$ vanishes.

The behavior in the disorder dominated quantum-critical regime
$t<x$, $r \ll t$ is more complicated.
For small momenta, $G_{\kappa \kappa'}$ is  approximated by 
$G_{\kappa \kappa'}\approx 
1/(2 \sqrt{r/t+\kappa^2+{\kappa'}^2-2 \kappa \kappa' \cos \phi})$,
where $\phi$ is the angle between $\vec{v}_{\k_H}$ and  
$\vec{v}_{\k_H+\Q}$. 
In leading order in $t/x$ for $b=0$ and $\kappa < 1 $
 Eqn.~\ref{boltz} is solved by
\begin{eqnarray}\label{pksimple}
p_\kappa^0 \sim t/x \int G_{\kappa \kappa'} d \kappa' \sim 
t/(2 x) \ln[ 1/(r/t+\kappa^2)].
\end{eqnarray}
where the prefactors and $\phi$ dependence in the argument of the $\log$
have been omitted.
The approximations leading to (\ref{pksimple}) are
 valid for $p^0_\kappa \ll 1$, therefore they break down in
an exponentially small regime
$r/t+\kappa^2 \lesssim c e^{-x/(2 t)}$, where $c$
 is a constant of order $1$.

\begin{figure}[t]
  \centering
 \epsfig{width=.8 \linewidth,file=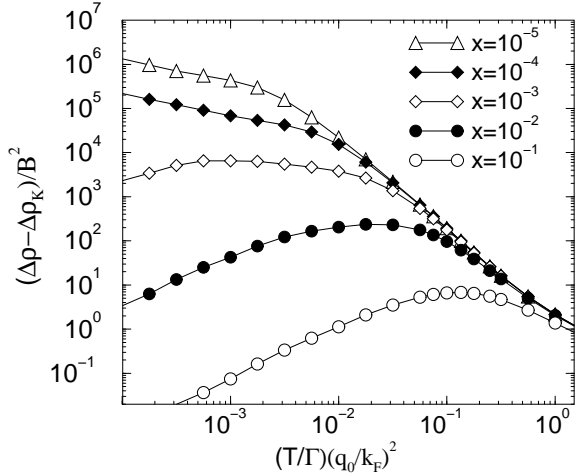}
\caption[]{Log-log plot of the temperature dependence of the
magnetoresistivity $\Delta \rho/b^2- \Delta \rho_K/b^2$ at the QCP in
a small magnetic field $b$ for model A, calculated from a numerical
solution of (\ref{boltz}). The ``Kohler's rule''-contribution $\Delta
\rho_K=\Delta \rho(t=0,b) (\rho(t=0,b=0)/\rho(t,b=0))$ has been
subtracted. In the dirty limit, $t<x$, one can indentify the
correction (\ref{magneto2}) with $\Delta \rho \sim b^2 t^2/(x^3
\sqrt{r}) \sim b^2 t^{5/4}/x^3$, $r \approx t^{3/2}$ at the QCP. The
magnetoresistivity in the clean regime, $x<t<\sqrt{x}$ is much larger with
$\Delta \rho \sim b^2/(\sqrt{x} t)$ (see Eqn.~(\ref{magneto3})).  In
the ultra-clean limit, $x<t^2$, we obtain $\Delta \rho \sim 1/t^2$ and
to leading order no effects of the QCP can be seen.}
\label{figDelRho}
\end{figure}

With all prefactors, which are calculated using (\ref{pdiffK}),  
for $e^{- x/(2 t)}\ll r < t <  x$,
the inelastic contribution to the
magnetoresistivity is given by
\begin{eqnarray}
\frac{\Delta \rho}{\rho_M}&\approx& b^2 \frac{t^2}{x^3 \sqrt{r}}
\biggl[ \frac{3 \pi^5 q_0^9/(2 v_F^4)}{(N_F (2 \pi)^3)^5} \sum_{i}
\oint_{i} d \k_{\|} \biggl[ \frac{v_{\k_{\|}}^3 \! + v_{\k_{\|}\pm
\Q_i}^3}{v_{\k_{\|}}^4 v_{\k_{\|}\pm \Q_i}^4}\nonumber \\ &&
\frac{((\vec{v}_{\k_{\|}}\!-\vec{v}_{\k_{\|} \pm \Q_i}) \vec{n})^2
((\vec{v}_{\k_{\|}} \times \vec{v}_{\k_{\|}\pm \Q_i})
\hat{\vec{B}})^2} {|\vec{v}_{\k_{\|}} \! \times \vec{v}_{\k_{\|}\pm
\Q_i}|} \biggr]\biggr], \label{magneto2}\end{eqnarray} where the
geometrical factors are accidentally exactly the same as in
(\ref{magneto1}). It is quite surprising, that even in the limit $r
\ll t$ the quadratic magnetoresistivity diverges proportional to
$1/\sqrt{r}$. What happens if a system is directly at the QCP, i.e. in
the case $r=0$?  In the scaling limit (\ref{scalinglimit}), the cutoff
$r/t$ in (\ref{pksimple}) is replaced by $e^{-x/(2 t)}$ and the
magnetoresistivity is proportional to $b^2 t^{1.5} e^{ x/(4 t)}/x^3
$. This exponentially large correction is probably not experimentally
observable. Either higher-order effects in the magnetic field will
become relevant (to be discussed in the next section) or one has to
take into account the temperature dependence of $r$.  The
low-temperature correction \cite{millis} $\Delta r \approx c' t^{3/2}$
can formally be neglected in the scaling limit and usually leads only
to small sub-leading corrections of the order of $\Delta r/t \approx
c' \sqrt{t}$ both in transport and thermodynamic quantities.  The
prefactor $c'$ is proportional to the spin-spin interaction $U$
(\ref{S}). The magnetoresistivity at the QCP in the dirty limit is a
remarkable exception of this rule: for $r \ll t \ll x$ and $c'
\sqrt{t}>e^{-x/(2 t)}$ the temperature dependence of $r$ determines
the $B^2$ correction to the resistivity and we expect $\Delta \rho
\sim b^2 t^{1.25}/(\sqrt{c'} x^3)$ (\ref{magneto2}) (
see Fig.~\ref{figDelRho}). An experimental
confirmation of this prediction would be an interesting way to
investigate the temperature dependence of $1/\xi^2$, a quantity which
can otherwise be measured only by high-precision inelastic neutron
scattering.

The solution of the Boltzmann equation (\ref{boltz1d})
in the regime $\max[x,\sqrt{r x}] \ll t<\sqrt{x}$ is dominated by large
momenta $\kappa, \kappa' \gg \max[ 1,\sqrt{r/t}]$. We therefore use
the asymptotic form of $G_{\kappa \kappa'}$ (\ref{Gasypt}) and rescale 
the argument of $p_{\kappa}$ by
$\kappa= q \sqrt{ \pi t/(6 x c_{\k_H} \sin^2 \phi)}/\sqrt{\alpha_H}$
and the argument of  $\tilde{p}_{\kappa'}$ by
 $\kappa'= q' \sqrt{ \pi t/(6 x c_{\k_H} \sin^2 \phi)} \sqrt{\alpha_H}$.
\begin{eqnarray}\label{boltztGTRx}
p_q   +\int  \frac{d q' (p_q+
\tilde{p}_{q'}-1) \sin^2 \phi}{(q^2/\alpha_H+{q'}^2 \alpha_H-
2 q q' \cos \phi)^{3/2}}&=& \frac{b \tilde{c}_{\k_H}}{t \sqrt{x}}
\partial_{q} p_q\nonumber \\
\tilde{p}_q+\int \frac{ d q'  (\tilde{p}_q+
p_{q'}-1)\sin^2 \phi}{(q^2 \alpha_H+{q'}^2/\alpha_H-
2 q q' \cos \phi)^{3/2}}&=&\frac{b \tilde{c}_{\k_H}}{t \sqrt{x}} \partial_{q} 
\tilde{p}_q 
\end{eqnarray}
with $\tilde{c}_{\k_H}=c'_{\k_H} |\sin \phi| \sqrt{6} \cos
\Theta_B/\sqrt{\pi c_{\k_H}}$. For small magnetic fields, $p_q$ decays
on the scale $1$ (both $\alpha_H$ and $\phi$ are finite for a generic
Fermi surface). This justifies our approximation to expand $G_{\kappa
\kappa'}$ for large momenta in the regime
$\sqrt{t/x}>\max[\sqrt{r}/t,1]$, i.e. for $t > \max[\sqrt{r x},x]$. In
the original units the width of $p_\k$ is approximately $\Delta k
\approx q_0 t/\sqrt{x}$ and by scaling we obtain the {\em linear}
resistivity shown in Eqn.(\ref{rholin}).

\begin{figure}[t]
  \centering
 \includegraphics[width=0.75  \linewidth]{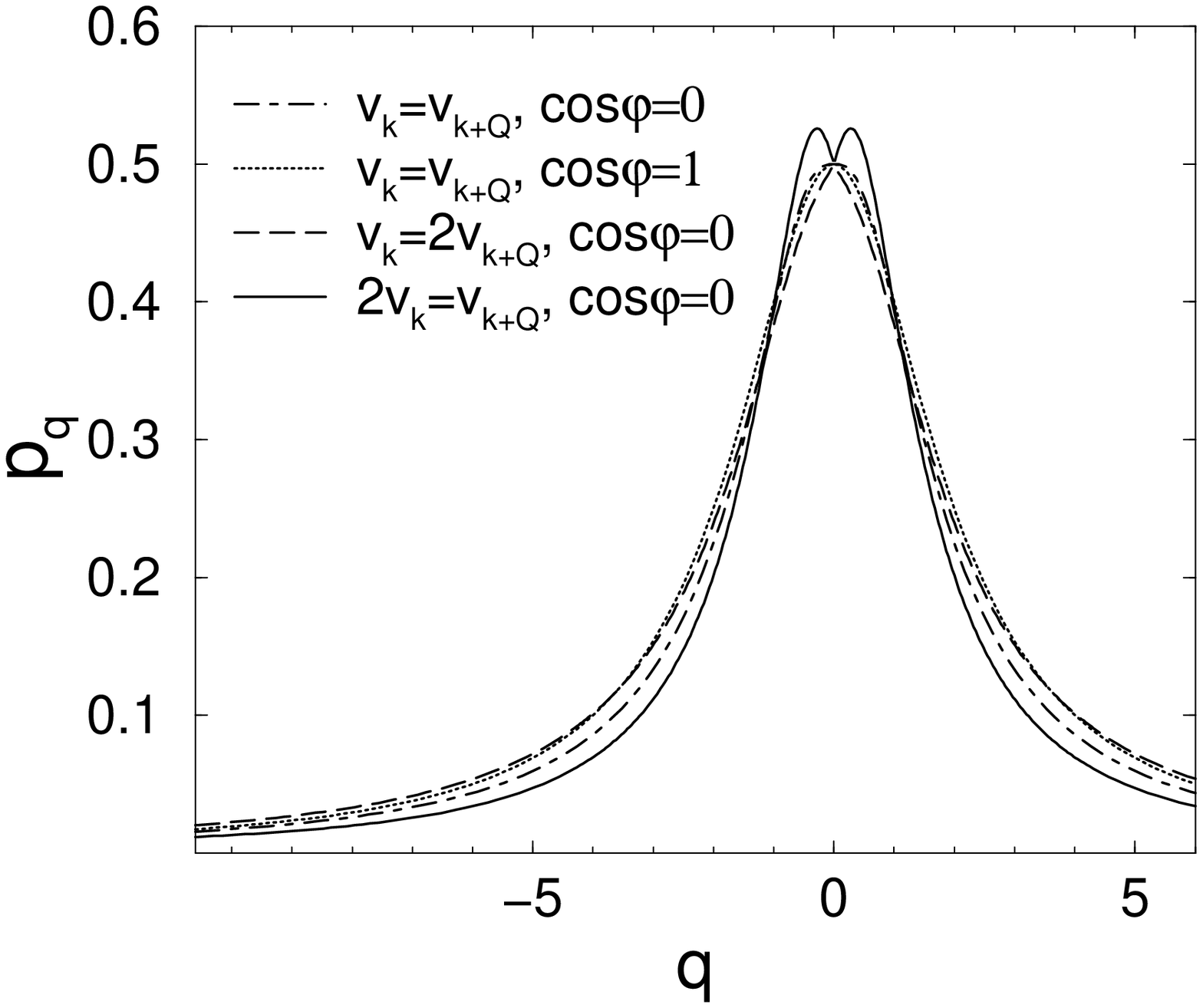}
\caption[]{Shape of $p_{q}$ in the regime $b=0$, $\max[\sqrt{x
r},x]<t<\sqrt{x}$ for various values of $\alpha_H=v_{\k \pm \Q}/v_\k$
and $\cos \phi=\vec{v}_{\k\pm \Q} \cdot \vec{v}_\k /(v_{\k\pm \Q}
v_\k)$.  The integral $S=\int p_{q}\approx \pi$ depends very weakly on
the parameters. $p_q$ with $k\propto q t/\sqrt{x}$ describes how
strong spin fluctuations equilibrate the distribution function
$\Phi_\k$ (\ref{pkdef}) and $\Phi_{\k \pm \Q}$ in a region of width
$t/\sqrt{x}$ around a hot line.}
\label{distQClean}
\end{figure}

While the kernel of (\ref{boltztGTRx}) is highly singular, the
solution of the integral equation for $b=0$ is nevertheless very
smooth: for large momenta $p_q$ decays proportional to $1/q^2$ and is
finite and smooth for $q\to 0$ with $p_0+\tilde{p}_0=1$ to cancel the
divergence of the kernel (see figure~\ref{distQClean}). The
perturbation theory in $b/(\sqrt{x} t))$ is analytic and well defined
(all eigenvalues of the matrix $\delta(q-q')(1+\int G_{q q'} d
q')+G_{q q'}$ are larger than $1$) and
we obtain for $b<\sqrt{x} t$, $\sqrt{x}> t > \max[\sqrt{r x},x]$:
\begin{eqnarray}\label{magneto3}
\frac{\Delta \rho}{\rho_M} \sim \frac{b^2}{\sqrt{x} t}
\end{eqnarray} 
Although it is difficult to calculate the precise prefactor
analytically, its structure is similar to the prefactor in
(\ref{magneto1}). In Fig.~\ref{figDelRho} the numerical solution of
the Boltzmann equation shows the crossover from the dirty limit
(\ref{magneto2}) to the clean limit (\ref{magneto3}). In the
ultra-clean limit, $x\ll t^2$ the distribution function varies on the
scale $k_F$ and Kohler's rule is valid in leading order and therefore
we expect $\Delta \rho \sim b^2/t^2$.

\subsection{Magnetoresistivity -- large fields}
The diverging prefactors of the $B^2$-corrections already indicate
that any finite magnetic field will strongly
influence the resistivity when the hot lines get sharper and sharper while
$B$ drives the quasi-particles over the Fermi surface.

For large fields, the main effect of the $b \partial_\k
p_k$-term in the Boltzmann equation (\ref{boltz1d}) is to keep 
$\partial_\k p_k$ small. Due to the boundary conditions 
$p_{k \to \pm \infty}\to 0$ and due to the fact that
the kernel $G_{k k'}$ decays for large momenta,
the amplitude of $p_k$ decreases for large magnetic fields and
$p_k$ is broadened. This corresponds to the fact that
magnetic field drives the quasi particles over the Fermi surface hence
smoothing the quasiparticle distribution.  In a strong enough magnetic
field we can therefore use perturbation theory in $p_k$ and only the
first term on the right-hand side of Eqn.  (\ref{pint}) survives. 
In the scaling regime, the resistivity saturates in a
large (orbital) magnetic field at a value which is precisely given by
the dirty-limit formula (\ref{rhoImp}).  In this respect, a magnetic
field and disorder have very similar effects, both smooth out the
quasi-particle distribution.

As the amplitude of $p_k$ decreases for large $b$, the main effect of
 a large magnetic field in the dirty limit $t<\max[x,\sqrt{r x}]$ is
 to suppress {\em sub-leading} corrections to the conductivity. The
 maximal change of $\int p_\kappa$ is calculated from (\ref{pint})
\begin{eqnarray}
\Delta \int p_\kappa(b \to \infty) &\approx& \frac{t}{x
\sqrt{\alpha_H} c_{\k_H}} \int \!\!\! \int_{-\infty}^{\infty} \!\!\! d
\kappa d \kappa' G_{\kappa \kappa'}
(p_{\kappa}^0+\tilde{p}_{\kappa'}^0) \nonumber \\ &=& (1+\alpha_H)
\int_{-\infty}^{\infty} \!\!\! d \kappa (p_{\kappa}^0)^2
\end{eqnarray}
where $p_{\kappa}^0=t/(x \sqrt{\alpha_H} c_{\k_H}) \int G_{\kappa
\kappa'} d \kappa'$ is the distribution function in the dirty limit in
leading order in $t/x$.

The effect of an intermediate field can be calculated by solving the
Boltzmann equation in a finite field  in perturbation theory in
$t/x$ by neglecting again the $G_{\kappa \kappa'} 
(p_{\kappa}+\tilde{p}_{\kappa'})$ terms.
\begin{eqnarray}
p_{\kappa}\approx \int_{\kappa}^\infty d \kappa' p^0_{\kappa'} 
\frac{e^{-(k'-k)/\kappa_b}}{\kappa_b}
\end{eqnarray}
with $\kappa_b= b/(x \sqrt{t}) (c'_{\k_H}/(\sqrt{\alpha_H} c_{\k_H}))$.
The main effect of the magnetic field is to smear the distribution function on
the scale $\kappa_b$.

In the Fermi liquid regime $t<\min[r,\sqrt{r x}]$ the scale of the
 distribution function
$p_{\kappa}^0 \sim (t/x)/(r/t+\kappa^2)$ is given by $\sqrt{r/t}$, accordingly
the small-field formula (\ref{magneto1})
 can be applied for  
$\kappa_b<\sqrt{r/t}$
or $b<\sqrt{r} x$. As discussed above, in a large magnetic field
$\Delta \rho$ defined in (\ref{delrho}) saturates at
\begin{eqnarray}\label{rhosat1}
\frac{\Delta \rho(b>\sqrt{r}x,t)}{\rho_M} \sim \frac{t^4}{x r^{3/2}}.
\end{eqnarray}
and the weak-field expansion (\ref{magneto1}) connects smoothly to
this regime. As above, we can estimate the range of validity of
(\ref{rhosat1}) by comparing it to neglected corrections of the
order of $\Delta \rho \sim b^2 \Delta \rho(0,t)/\rho_0^2 \sim 
b^2 t^2/(\sqrt{r} x^2)$. Therefore (\ref{rhosat1}) is valid for $b<t
\sqrt{x/r}$, $t<\min[x,\sqrt{r x}]$.

The dirty quantum-critical regime $r<t<x$ has a richer structure, as
there are two scales in the distribution function $p_\kappa^0$
(\ref{pksimple}): $1$ and $\max[\sqrt{r/t},e^{- x/(2 t)}]$. For
$1>\kappa_b>\max[\sqrt{r/t},e^{- x/(4 t)}]$ the magnetic field, or more
precisely $\kappa_b^2$, takes the role of $r/t$ in (\ref{magneto2})
and provides a cutoff to the logarithmic divergence of $p_\kappa^0$
(\ref{pksimple}).  We therefore expect for $r<t<x$, $\max[x \sqrt{t}
e^{- x/(2 t)},\sqrt{r} x]< b<x \sqrt{t}$
\begin{eqnarray}\label{linb}
\frac{\Delta \rho(b,t)}{\rho_M} \sim b \frac{t^2}{x^2}
\end{eqnarray}
and the magnetoresistivity will saturate for $b>x \sqrt{t}$ at
\begin{eqnarray}\label{rhosat2}
\frac{\Delta \rho(b,t)}{\rho_M} \sim \frac{t^{5/2}}{x}
\end{eqnarray}
This has to be compared to the largest sub-leading correction of the form
$b^2 t^{1.5}/x^{2}$ (estimated from Kohler's rule). Therefore equation
 (\ref{rhosat2}) is valid for $b<\sqrt{t x}$.

All the above discussed effects in the dirty limit are very small on 
an absolute scale
because the disorder has already smeared out most of the features of the
distribution function and can only be observed because it is easy to
measure the difference $\rho(T,B)-\rho(T)$ with high precision.
 Much stronger effects are to be expected in
the cleaner regime $\max[x,\sqrt{r x}]<t<\sqrt{x}$, where the zero-field
resistivity is linear in temperature.

\begin{figure}[t]
  \centering
 \includegraphics[width=0.75 \linewidth]{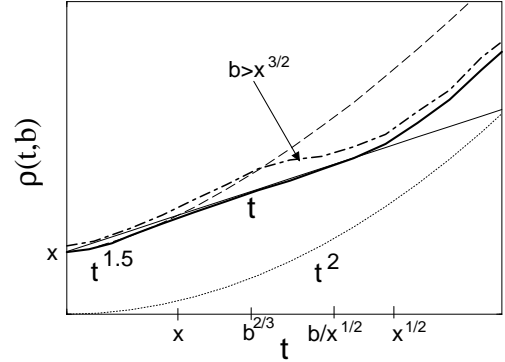}
\caption[]{Schematic plot of the temperature dependence of the
resistivity.  In an ultra-clean system, the resistivity would rise
proportional to $t^2$ (thin dotted line). In a system with a small
amount $x$ of disorder the resistivity at the QCP (thick solid line)
at zero magnetic field rises with $t^{3/2}$ at very low temperatures
$t<x$, but shows a linear behavior $\Delta \rho \sim \sqrt{x} t$ for
$x<t<\sqrt{x}$. If the system is tuned to the quantum-critical point in
a finite field, a behavior similar to the dot-dashed line is expected:
 $\Delta \rho \propto t^{3/2}$ should be observed over
a larger range $t<b^{2/3}$. In an intermediate regime
$b^{2/3}<t<b/\sqrt{x}$ the resistivity depends weakly on
temperature (the thin dashed line corresponds to $\Delta \rho \propto
t^{3/2}$). Note that a different behavior is expected if a finite magnetic
field is applied to a compound which is quantum critical 
for $b=0$ as $b$ drives the system away from the QCP.}\label{figrhoMagSchema}
\end{figure}
For $b<\sqrt{x} t$, the effect of the magnetic field can be treated
in perturbation theory and (\ref{magneto3}) is valid. 
To analyze the regime
$\sqrt{x} t < b$ it is useful to rescale the momenta in
(\ref{boltztGTRx}) using
 $q=-b \tilde{c}_{\k_H}/(t \sqrt{x}) k \sqrt{\alpha_H}$ so
that the Boltzmann equation reads
\begin{eqnarray}
\sqrt{\alpha_H} p_k + \frac{t^2 x}{b^2 \tilde{c}_{\k_H}^2} \int
\frac{d k' (p_k+ \tilde{p}_{k'}-1) \sin^2 \phi}{(k^2+{k'}^2 - 2 k k'
\cos \phi)^{3/2}}&=& -\partial_{k} p_k \nonumber \\ \label{boltzBL}
\end{eqnarray}
The corresponding equation for $\tilde{p}_{k'}$ is obtained by
 replacing $\alpha_H$ by $1/\alpha_H$.  The kernel of the integral
 equation is strongly divergent $\int G_{k k'} d k' \propto
 1/k^2$. The contribution from the integral is, however, small for
 large magnetic fields and sufficiently large momenta $k>k_0$ (a
 precise estimate is given below) and (\ref{boltzBL}) is solved by
 $p_{k \gtrless 0}=c_{\pm} e^{- k \sqrt{\alpha_H}}$.  For $k<0$ the
 prefactor $c_{-}$ has to vanish due to the boundary condition
 $p_{|k|\to \infty}=0$. For $k \to 0$ the divergence of the kernel
 dominates the integral equation and forces the solution to take the
 form $p_{k \to 0}+\tilde{p}_{k'\to 0}=1$. 
For   $\alpha_H=1$  one therefore expects \cite{alphaH1}
 $p_{k \to 0}=\tilde{p}_{k'\to 0}=1/2$.
 Generally, $p_{k \to 0}$ is a
 number close to $0.5$ depending slightly on $\alpha_H, \phi$ and
 $b$. For a large magnetic field, the rise of $p_k$ close to $k=0$ is
 independent of $\alpha_H$ and $p_{k\to 0}(b\gg \sqrt{x}
 t)=\tilde{p}_{k\to 0}(b\gg \sqrt{x} t)=1/2$.  The prefactor of the
 exponential decay $c_+$ for $k>k_0$ is therefore $1/2$ and $\int p_k
 d k=1/(2 \sqrt{\alpha_H})$ or, after rescaling, $\int p_{\kappa}
 d\kappa = b/(x \sqrt{t}) |\!\cos \Theta_B| (c'_{\k_H}/(2 \alpha_H
 c_{\k_H}))$.   In this
 regime $\sqrt{x} t < b <\min[t^{3/2}, t^2/\sqrt{r}]$, $\max[x,\sqrt{r
 x}]<t<\sqrt{x}$, the rise of the resistivity $\rho-\rho_0$
 is linear in $|\vec{B}|$ and {\em independent} of
 temperature and disorder (the precise range of validity is calculated
 below).
\begin{eqnarray}\label{rhoBlin}
\frac{\rho(b,t)-\rho_0(b)}{\rho_M}&\approx& b \biggl[\frac{ 3 \pi q_0^3/2}{v_F^4 ((2 \pi)^3 N_F)^2}
\sum_{i} \oint_{i} 
d \k_{\|} \biggl[ \\
&&  \frac{((\vec{v}_{\k_{\|}}\!-\vec{v}_{\k_{\|} \pm \Q_i}) 
\vec{n})^2  |(\vec{v}_{\k_{\|}} \times \vec{v}_{\k_{\|}\pm \Q_i}) 
\hat{\vec{B}}|}
{|\vec{v}_{\k_{\|}} \! \times \vec{v}_{\k_{\|}\pm \Q_i}|} 
\biggr]\biggr] \nonumber
\end{eqnarray}
For higher magnetic fields the formula given above will break down,
because (\ref{boltztGTRx}) is valid only for large momenta 
$\kappa>\max[1,\sqrt{r/t}]$ or, equivalently, (\ref{boltzBL}) is
valid for $k>\max[x \sqrt{t}/b,\sqrt{r} x/b]$. To estimate the range of
validity of (\ref{rhoBlin}), we  solve  (\ref{boltzBL}) more
precisely for $k<0$ in a regime, where $p_{k}$
is still small:
\begin{eqnarray}
p_k \approx  \frac{t^2 x}{b^2 \tilde{c}_{\k_H}^2} \int_{-\infty}^k 
d k' e^{(k'-k) \sqrt{\alpha}} \frac{2}{k^2} \approx  
\frac{t^2 x}{b^2 \tilde{c}_{\k_H}^2} \frac{2}{k}
\end{eqnarray}
$p_k$ is large for $k \lesssim k_0 \approx t^2 x/b^2$ and
Eqn.  (\ref{rhoBlin}) is therefore valid for $k_0>\max[\sqrt{t}
x/b,\sqrt{r} x/b]$ or $b<\min[t^{3/2}, t^2/\sqrt{r}]$. For larger
values of $b$ the kernel is less singular and $p_k$ stays small for
all momenta.  Accordingly, the resistivity is dominated by the first
term on the right-hand side of Eqn. (\ref{pint}) and the
magnetoresistivity saturates at a value, which is given by the
dirty-limit formula (\ref{rhoImp}):
\begin{eqnarray}
\frac{\rho(b,t)-\rho_0(b)}{\rho_M}&\approx&
t^{3/2} h\!\left(\frac{t}{r}\right)  \frac{3 q_0^3}{8 \pi^4 (v_F N_F)^2 }
\nonumber \\  \label{rhoImpLargeB}
&& \times \sum_i 
\oint_{i} d \k_{\|} \frac{((\vec{v}_{\k_{\|}}-\vec{v}_{\k_{\|} \pm \Q_i}) 
\vec{n})^2}
{|\vec{v}_{\k_{\|}} \times \vec{v}_{\k_{\|}\pm \Q_i}|} 
 \\
h\left(\frac{t}{r}\right)
&\approx& \left\{ \begin{array}{ll}
\frac{\pi \zeta(3/2) \Gamma(5/2)}{2 \sqrt{2}} 
 &, t >r  \\[.3cm]
\frac{\pi^3}{12}\sqrt{t/r}
&, t< r  
\end{array} \right.\nonumber
\end{eqnarray}
This expression is valid for $\min[t^{3/2}, t^2/\sqrt{r}]< b
<\min[t^{1/4} x^{3/4},\sqrt{t} x^{3/4}/r^{1/4}]$, where the upper
limit estimates the regime where non-universal contributions following
Kohler's rule dominate. In Fig.~\ref{figrhoMagSchema} the temperature
dependence of the large-field magnetoresistivity is shown
schematically.

\section{Discussion} \label{discussionSection}
\subsection{Overview over magnetoresistivity}\label{overview}
Before we discuss the experimental relevance of our calculations in
detail, we collect the results for the magnetoresistivity in the
scaling limit (\ref{scalinglimit}).  Most of the prefactors, which are
omitted here, can be found in the previous sections. Only {\em orbital}
effects of the magnetic field are taken into account, a discussion of
the interplay of spin and orbital effects is given in section
\ref{sectionSpinOrbit}.  $t$ and $b$ defined in (\ref{defConst}) are
the dimensionless temperature and magnetic field, respectively. $x$ is
proportional to the strength of disorder and can crudely be identified
with the inverse of the residual resistivity ratio and $r$ measures
the distance from the QCP.

In the disorder dominated regime the absolute size of effects of an
orbital magnetic field are quite small. Close to the QCP, for $r \ll
t<x$, the prefactor of the $b^2$-term is, however, extremely large and
for very small magnetic fields a crossover to a linear field
dependence of the magnetoresistivity is predicted. For $\Delta
\rho=\rho(b,t)-\rho(b,0)$ we obtain
\begin{eqnarray}
\frac{\Delta \rho}{\rho_M} \sim t^{3/2}- \frac{t^{5/2}}{x} + \left\{ 
\begin{array}{ll}
\displaystyle \frac{ b^2 t^{2}}{x^2 g_{x, r, t}}\! & , b<g_{x, r, t}
\\[1 em] \displaystyle b \frac{t^2}{x^2} & , g_{x, r, t}<b<x \sqrt{t}
\\[1 em] \displaystyle \frac{t^{5/2}}{x} & , x \sqrt{t}<b<\sqrt{t x}\!
\end{array}\right.
\end{eqnarray}
with $g_{x, r, t} \approx \max[x \sqrt{r},x \sqrt{t} e^{- x/(4 t)}]$.
For very small $g_{x, r, t}$ this expression can be used to measure
the temperature dependence of the correlation
length $\xi$ using $r \propto 1/\xi^2$.

In the disorder-dominated Fermi liquid regime $t<\min[r,\sqrt{r
x}]<1$, the temperature dependent part of the magnetoresistivity is
still enhanced, but it is probably very difficult to extract the $t^4$
contribution experimentally.
\begin{eqnarray}
\frac{\Delta \rho}{\rho_M} \sim \frac{t^{2}}{\sqrt{r}}- \frac{t^4}{x r^{3/2}}
 + \left\{ 
\begin{array}{ll}\!
 \displaystyle \frac{b^2 t^4}{x^3 r^{5/2}}\! & , b<\sqrt{r} x, t>r \sqrt{x}
\!\!\! \\[1 em]
\! \displaystyle \frac{t^4}{x r^{3/2}}  & ,  \sqrt{r} x<b<t \displaystyle
\sqrt{\frac{x}{r}}
\end{array}\right.
\end{eqnarray}

The largest effects can be observed for rather  clean systems
with $\max[x,\sqrt{r x}]<t<\sqrt{x}$, which show a linear resistivity for
$b=0$:
\begin{eqnarray} \label{rhoTB}
\frac{\Delta \rho}{\rho_M} \sim  \left\{ 
\begin{array}{ll}
 \displaystyle \sqrt{x} t+ \frac{b^2}{\sqrt{x} t} & , b<\sqrt{x} t
\\[1 em] \displaystyle b & , \sqrt{x} t
<b<\min[t^{3/2},\displaystyle\frac{t^2}{\sqrt{r}}] \!\! \\[1 em]
\displaystyle \min[t^{3/2},\frac{t^2}{\sqrt{r}}] \! & , \min[t^{3/2},
\displaystyle\frac{t^2}{\sqrt{r}}]<b< \\ & \qquad x^{3/4}
\min[t^{1/4},\displaystyle \frac{\sqrt{t}}{r^{1/4}}]
\end{array}\right.
\end{eqnarray}

\subsection{Experimental situation for $B=0$}

According to our calculation, in $d=3$, spin-fluctuation theory
predicts that the resistivity in dirty systems rises with $T^{3/2}$,
while in cleaner compounds a large regime with a linear resistivity is
expected.  This behavior is actually seen in a large number, but not
all, AFM quantum-critical systems.  A resistivity proportional to
$T^{1.5}$ has been reported in CeCu$_2$Si$_2$
\cite{gegenwart}, CeNiGa$_2$ \cite{hauser}, CeCu$_{6-x}$Ag$_x$
\cite{heuser} CeNi$_2$Ge$_2$ \cite{grosche}, CePd$_2$Si$_2$, CeIn$_3$
\cite{julian} or CeCu$_5$Au \cite{wilhelm}. All these systems are
dirty in the following sense: in the regime, where the exponent $1.5$
has been observed, the rise of the resistivity $\Delta \rho$ is small
compared to the residual resistivity.

\begin{figure}[t]
  \centering
 \includegraphics[width=\linewidth]{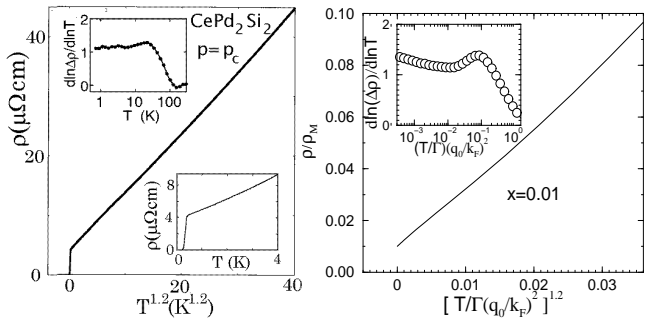}
\caption[]{Resistivity as a function of $T^{1.2}$ in CePd$_2$Si$_2$
taken from \cite{julian} (left figure) compared to our calculation
within model A defined in appendix~\ref{appendixModelA} for $x=0.01$
(right figure).  The insets show the corresponding logarithmic
derivative of $\rho(T)-\rho(0)$. If one compares the insets one has to
take into account that the low-temperature behavior of the logarithmic
derivative of $\Delta \rho$ strongly depends on the exact
determination of $\rho_0$ (see inset of Fig.~\ref{r01fig}).  Below
$\approx 400$mK CePd$_2$Si$_2$ is superconducting (lower inset).  Note
the offset of the line $T=0$ in both plots.}
\label{figExperiment}
\end{figure}
For a small number of the above mentioned compounds, cleaner samples
exist.  For these high-purity single crystals, the resistivity seems
to rise with an exponent smaller than $1.5$. The cleaner the samples
are, the closer the exponent is to $1$. This qualitative trend
coincides with the predictions of our calculations
(Fig.~\ref{figExponent}). Examples for this are CePd$_2$Si$_2$,
CeNi$_2$Ge$_2$ \cite{mathur,julian,grosche} where samples with RRR of
the order of 100 exist. In clean samples of CeIn$_3$, exponents
smaller than $1.5$ are observed over some temperature range, but at
lowest temperatures the exponent crosses over to $1.5$, as expected
from theory.  In Fig.~\ref{figExperiment} the resistivity of a quite
clean sample of CePd$_2$Si$_2$ \cite{julian} is compared to a typical
solution of the Boltzmann equation.  It is important to point out,
that not only a similar effective exponent shows up in both theory and
experiment, but that it is observed over a similar range $0.1 \rho_0 <
\Delta \rho(T) <10 \rho_0$. Only the effective scale $\Gamma/q_0^2$
seems to be approximately a factor $10$ too large compared to
$T_K/k_F^2$, but a quantitative comparison of these high-energy
scales is certainly very dubious within our perturbative treatment of
a strongly correlated system.  Furthermore, a quantitative comparison
would require a more realistic model based on some knowledge of the
band structure and the actual width of the spin-fluctuation.

The theory, presented in this paper, is certainly not applicable to
strongly disordered systems, where the nature of the quantum-critical
transition is changed due to disorder. It does not explain the linear
temperature dependence of the resistivity in CeCu$_{6-x}$Au$_x$ due to
two reasons. The first is, that CeCu$_{6-x}$Au$_x$ is relatively dirty
at the QCP ($x=0.1$).  More important is the unusual
nature of the spin-fluctuations, which are strongly anisotropic
\cite{stockert}, and show an unusual frequency dependence
\cite{schroeder}.

\begin{figure}[t]
  \centering
 \includegraphics[width=0.7 \linewidth]{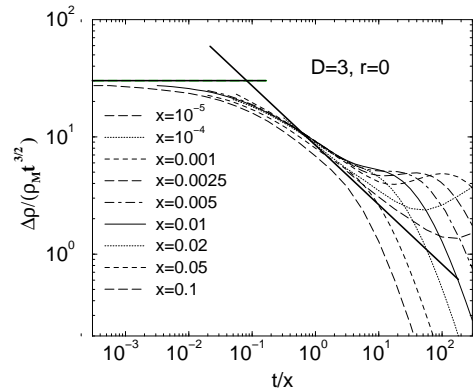}
\caption[]{Scaling plot of $f(t/x,0)=\Delta \rho(t/x)/t^{3/2}$ 
at the QCP ($r=0$) using a numerical solution of the Boltzmann equations
for various amounts of disorder $x$.
The straight solid lines show the limiting cases $\Delta \rho \propto t^{3/2}$
for $t \ll x$ and $\Delta \rho \propto t$ for $x \ll t \ll \sqrt{x}$.
Deviations from scaling are large for $t/x>1/\sqrt{x}$.}
\label{figScaling}
\end{figure}
The most precise test of the spin-fluctuation model would be to verify
(or falsify) our scaling prediction (\ref{scaling1}) by comparing
several very clean quantum-critical samples with varying residual
resistivity ratios.  An example for the expected deviations from
scaling is shown in Fig.~\ref{figScaling}.

A somewhat unexpected result of our calculation is that in clean
systems away from the QCP, a Fermi liquid form of the resistivity can
be observed only at very low temperatures $t<\sqrt{r x}$. This has to
be compared to thermodynamic quantities, where FL behavior sets in at
a considerably higher scale $t<r$. In a clean system with a RRR over
$100$, the crossover scale in transport quantities can easily be
e.g. a factor $5$ lower than in thermodynamic quantities.  This might
explain claims based on transport measurements
that the non-Fermi liquid behavior in these clean
systems extends over a finite region away from the QCP
\cite{mathur,grosche}.

\subsection{Spin- and orbital effects of $B$}\label{sectionSpinOrbit}
A magnetic field influences the resistivity near a quantum critical
point in two quite independent ways: by spin- and orbital effects.
The spin contribution typically suppresses the antiferromagnetic
order. In the context of this paper, we will call these phenomena
``spin effects'' despite the fact, that in most cases orbital degrees
of freedom are involved. On the non-magnetic side of the phase diagram
and for $B$ fields, which are not too large, the leading order
spin-effects can supposedly be described by the magnetic field
dependence of $r$, the variable which describes the distance to the
QCP. For small fields, one expects in an AFM metal above the upper
critical dimension $\Delta r \propto B^2$.  A controlled study of the
effects of a finite magnetic field close to the AFM QCP in a metal
does not exist to our knowledge, however, Ioffe and Millis
\cite{ioffeSusz} have carefully investigated the uniform
susceptibility without finding unexpected anomalies. A large magnetic
field can induce other (first-order) transitions or change the nature
of the magnetic fluctuations completely.

 The focus of this paper are ``orbital effects''. We consider a
regime where interference effects don't play any role and have
restricted our attention to the classical $\vec{v}\times \vec{B}$
force on the electrons. This approximation is valid in the scaling
limit (\ref{scalinglimit}), where $\w_c \tau \ll 1$ and the disorder is so
weak that weak-localization corrections set in at much lower
temperatures \cite{altshuler}.

In the scaling limit (\ref{scalinglimit}), i.e. for clean enough
systems at very low temperatures in small magnetic fields close to the
QCP, spin effects can be neglected. For $b \propto x^{3/2}$
the correction $\Delta r \propto b^2$ is small compared to $x$ for
sufficiently small $x$.  In realistic heavy Fermion systems, it is,
however, probably not possible to produce a sample which is clean
enough so that {\em  only} orbital effects show up in the interesting
regime. The reason is that on the one hand the Lorentz force $\vec{v}
\times \vec{B}$ is suppressed due to the large mass $m^*$ of the heavy
quasiparticles and on the other hand spin-effects are enhanced due to
the large susceptibility $\chi \approx 1/T_K \propto m^*/m_0$.

A reliable estimate of these effects is difficult as the details of
the interplay of the Kondo effect and the RKKY interaction are poorly
understood\cite{colemanSUSY}.  For a crude estimate, we assume that
all high-energy scales, including scattering rates and renormalized
coupling constants can be approximated by $T_K$. Orbital effects are
important for $b>x^{3/2}$ or $\mu_B B > x^{3/2} k_B T_K m^*/m_0$ and
spin effects can be neglected for $\Delta r<x$ or $\mu_B B < \sqrt{x}
k_B T_K \sqrt{m_0/m^*}$.  If this is qualitatively correct, only an
extremely clean sample with $x<(m_0/m^*)^{3/2}$, i.e. with a RRR
larger than $(m^*/m_0)^{3/2}$, shows orbital effects without the
influence of spin effects. This implies that spin-effects are present
in most cases, because $m^*/m_0$ is quite large in those clean
stoichiometric systems where the quantum critical point is reached by
pressure.

Nevertheless, the orbital effects
have a very clear signature even in the presence of large
spin-effects, especially in suitably designed experiments. The
necessary magnetic fields are  not very large. In our units,
Shubnikov-de Haas oscillations are expected for $b > x$ at low
temperatures while strong effects in the magnetoresistivity in clean
samples set in for $b>x^{3/2}$, i.e. for magnetic fields which can
easily be an order of magnitude smaller.

The Hall constant is obviously not influenced by O($B^2$) corrections
due to the suppression of the AFM and  so $R_H$ is
not affected by the above mentioned problems. Therefore it should be
very interesting to investigate the strong sample dependence of the
Hall effect predicted in section {\ref{hallSection}}.  Unfortunately,
only extremely clean samples will show a strong temperature dependence
(see Fig.~\ref{figHall}) \cite{skew}. 

For a detailed study of the magnetoresistivity in clean systems, it is
necessary to minimize spin effects.  In a system like CePd$_2$Si$_2$
the effective distance from the QCP can be tuned by pressure.
Therefore, one can select various pressures on the antiferromagnetic
side of the phase diagram close to the QCP and suppress the remaining
magnetic order by a suitable magnetic field. For such a set-up the
spin effects are compensated by a change of pressure to a large extent
especially in systems with Ising symmetry.  For not too strong fields
the field-tuned quantum-critical transition should stay second order
and one can compare several quantum-critical systems ($r=0$) with the
same amount of disorder $x$ but different magnetic fields. Our
calculation predicts, that in a magnetic field $b>x^{3/2}$, the
low-temperature resistivity rises proportional to $t^{1.5}$. A
schematic plot of the expected behavior is shown in
Fig.~\ref{figrhoMagSchema}.  Ideally, one can test the scaling
predictions by comparing samples of different quality.  In an
anisotropic system one should choose a direction of the magnetic field
perpendicular to the easy axis and to one of the ordering vectors
$\Q$, because only the component of $B$ parallel to the hot lines
enters the scaling form of the resistivity. (This effect is described
by the $(\vec{v}_{\k_{\|}} \times \vec{v}_{\k_{\|}\pm \Q_i})
\hat{\vec{B}}$ factors in our results, e.g. in (\ref{rhoBlin}).)

A simple way to distinguish spin- and orbit effect is the sign of the
magnetoresistivity. A suppression of the AFM fluctuations on the
paramagnetic side of the phase diagram will reduce the amount of
scattering and the resistivity will go down. Due to the orbital
effects discussed in the previous section resistivity {\em increases}
because $b$ smears out the quasi particle distribution, therefore
minimizing the effect that cold regions short-circuit the hot spots.
This can lead to a sign change of $\Delta
\rho=(\rho(B,T)-\rho(0,T))-(\rho(B,0)-\rho(0,0))$ either as a function
of disorder or strength of the magnetic field. Such an observation
could help to separate the two effects quantitatively.

The interplay of the spin- and orbit effects can also lead to less
trivial effects. In the dirty limit close to the QCP e.g. the
magnetoresistivity for small magnetic fields (\ref{magneto2}) is
proportional to $1/\sqrt{r}$, $\Delta \rho \sim b^2 t^2/(x^3
\sqrt{r})$. If spin effects suppress the AFM fluctuations with $r
\approx c b^2$, one can expect a regime where the magnetoresistivity
rises {\em linearly} with $B$ for rather small fields ($c
b^2<r(b=0,t)$), $\Delta \rho \propto b t^2/(\sqrt{c} x^3)$. It is
probably not very easy to distinguish this effect from the linear $b$
dependence due to orbital effects discussed in (\ref{linb}) with
$\Delta \rho \sim b t^2/x^2$ for $b>x \sqrt{r(b=0)}$.

In high purity single crystals of CeNi$_2$Ge$_2$ Grosche {\it et al.}
\cite{grosche} have reported measurements of the resistivity in
magnetic fields of a few Tesla. As explained above, a direct
comparison of their findings with our calculations is difficult due to
the interplay of orbital and spin effects. However, an important
qualitative trend of their results can easily be explained by our
approach.  In finite magnetic field, they observe a clear $T^2$
dependence below a temperature $T^*(B)$. Obviously, the magnetic field
drives the system away from the quantum-critical point. Remarkably, it
seems that the magnetic field is much more effective in this respect
than pressure, which also suppresses the magnetic order. For
samples tuned by pressure towards the paramagnetic side of the phase
diagram, a clear $T^2$-rise of the resistivity could not be observed.
This qualitative effect can be understood within our approach.
In the case of pressure tuning, we expect Fermi liquid in a very clean 
sample  only for very low temperatures $t<\sqrt{r x}$. In a sufficiently
strong magnetic
field, the crossover to $T^2$ should be visible at a much higher 
temperature $t<r^{1/4} \sqrt{b}$ in agreement with the experimental trends.
If we assume that $r \propto b^2$ and use  the above discussed crude
estimates of the prefactors, we expect the crossover temperature $T^*$
to rise {\em linear} in $B$ with $k_B T^* \approx \mu_B B (m_0/m^*)^{1/4}$.
I want to emphasize that this estimate is unreliable and should be replaced
by a suitable strong-coupling theory of heavy-Fermion systems. Nevertheless,
the experimental results \cite{grosche} 
seem to be consistent with this simple-minded
estimate of scales.

\section{Conclusion} \label{conclusions}

In all situations where two scattering mechanisms compete, one
anisotropic and affecting only parts of the Fermi surface, the other
isotropic, the transport is quite subtle and the interplay of the two
mechanisms, $A$ and $B$, can lead to qualitatively new effects.  The
reason for this is very simple: the effectiveness of scattering
mechanism $A$ in the presence of a current is strongly influenced by
the momentum dependence of the out-of-equilibrium distribution of the
quasi particles, which is affected by mechanism $B$.  Therefore,
we expect that $\rho_{AB} \neq \rho_A+\rho_B$ even in the
weak-coupling limit of a semi-classical theory described by a Boltzmann
equation ($\rho_X$ is the resistivity in the presence of scattering
mechanism $X$).

In this paper we have analyzed such a situation in detail. Near the
AFM quantum-critical point of slightly disordered metals, weak
isotropic impurity scattering competes with the strongly anisotropic
scattering from spin fluctuations, which is most effective along
``hot'' lines on the Fermi surface. The observation that even in
relatively clean single crystals the effective resistivity exponent
$\alpha$, $\Delta \rho \propto T^{\alpha}$, depends on the amount of
disorder in the system \cite{julian,mathur,grosche}, is an important
indication that the interplay of scattering mechanisms is relevant in
these systems.  Our results for $B=0$ explain the experimental trends:
$\Delta \rho \propto T^{1.5}$ for dirty systems and $\Delta \rho
\propto T$ in high purity samples.  A precise test of the
spin-fluctuation picture which underlies our calculation would be to
check our scaling predictions using a comparison of several
high-purity samples with different residual resistivities $\rho_0$
directly at the QCP: $\Delta \rho \approx T^{1/5} f(T/\rho_0)$ with
$f(y \to 0)=const.$ and $f(y \to \infty) \propto 1/\sqrt{y}$.

We furthermore propose to use orbital effects in a large magnetic
field to study anisotropic scattering mechanisms in detail.
Interesting nonlinear effects are expected if the magnetic field is
strong enough to push quasi particles within the scattering time $\tau_\k$
over any sharp structure with width $\Delta k$ on the Fermi surface,
i.e. for $e (\vec{v} \times \vec{B}) \tau_{\k} > \Delta k$.  In the
limit of very small but finite disorder, we were able to calculate the
non-linear magnetoresistivity analytically in all relevant limits. The
main effect  of a magnetic field 
is that sharp features in the quasiparticle distribution are smeared out.
In this respect both an increase of disorder and a magnetic fields lead
to similar effects. In a sufficiently strong 
magnetic fields, we therefore expect a $T^{1.5}$ rise of the
resistivity at the QCP and a $T^2$ law on the Fermi-liquid side of the
phase diagram even in samples where the resistivity is close to linear in
the absence of the magnetic field. Precise scaling predictions can be
used to check this mechanism experimentally.

We hope that the results of this paper establish a set of predictions
 which can be used to investigate mainly one question:
Can the non-Fermi liquid behavior in clean systems be explained 
by a nearly antiferromagnetic Fermi liquid? At the moment this question
is still open. Our interpretation of the data e.g. of CePd$_2$Si$_2$
favors a SDW interpretation. Certainly it is necessary to look also
at other quantities. For example in CePd$_2$Si$_2$ the Ne\'el temperature
seems to rise linearly with $p-p_c$ while in spin-fluctuation theory an
exponent $2/3$ is expected.

It is important to stress that at least in some systems close to an
AFM QCP a description within the three-dimensional SDW scenario seems
to be impossible.  The best studied example for this is
CeCu$_{5.9}$Au$_{0.1}$ \cite{loehneysen} where the spin-fluctuations
spectrum shows both an anomalous momentum \cite{stockert} and energy
\cite{schroeder} dependence.  The linear temperature dependence of the
resistivity in CeCu$_{5.9}$Au$_{0.1}$ has a different origin than the
one discussed in this paper. As the residual resistivity is relatively
high, the resistivity should rise with $T^{1.5}$ within the $3d$
models discussed in this paper (a linear resistivity could result from
$2d$ critical fluctuations coupled to $3d$ electrons \cite{stockert}).
Interestingly, an exponent $1.5$ has been reported for the quantum
critical resistivity of CeCu$_{6-x}$Ag$_{x}$ \cite{heuser} and
pressure tuned CeCu$_5$Au \cite{wilhelm}, which might indicate that
small changes can induce a more conventional SDW-transition in this
compound.  CeCu$_{5.9}$Au$_{0.1}$ is by now the most likely and best
studied candidate \cite{schroeder,coleman,si} for a truly
unconventional QCP (YbRh$_2$Si$_2$ \cite{trovalli} and
U$_{0.2}$Y$_{0.8}$Pd$_2$Al$_3$ \cite{maple} have very similar
thermodynamic and transport properties). The reported scaling of the
susceptibility in CeCu$_{5.9}$Au$_{0.1}$ \cite{schroeder} seems to
contradict any simple SDW models \cite{stockert}. An interesting
question is whether the rather strong disorder is responsible for the
anomalous behavior or whether a pure interaction effect destroys the
Fermi liquid close to the QCP.

\section{Acknowledgement}

I would like to thank N.~Andrei, R.~Chitra, A.V. Chubukov, P.~Coleman,
P.~Gegenwart, M.~Grosche, S.~Julian, G.~Kotliar, E.~Lange,
H.~v.~L\"ohneysen, A.~Millis, C.~Pfleiderer, A.E.~Ruckenstein and
P.~W\"olfle for discussions. This work was supported by the
A.~v.~Humboldt foundation and the Center for Materials Theory at
Rutgers University.

\begin{appendix}

\section{Approximate variational solution of the Boltzmann equations}
\label{appVar}
In the absence of the magnetic field the Boltzmann equation is
equivalent to the following variational problem\cite{ziman,hlubina}
which allows to calculate the qualitative behavior of the resistivity
in a very simple and efficient way. Following Hlubina and
Rice \cite{hlubina}, we define a functional $\rho$ of $\Phi_{\vec{k}}$
\begin{eqnarray}
\rho[\Phi_{\vec{k}}]&=& \frac{\hbar}{4 e^2}\frac{\oint\!\!
 \oint
\frac{d \vec{k}d \vec{k}`}{v_{\vec{k}}v_{\vec{k}'}}
 F_{\vec{k}\vec{k}'} (\Phi_{\vec{k}}-\Phi_{\vec{k}`})^2}
{\left(\oint\frac{d \vec{k}}{v_{\vec{k}}} (\vec{v}_{\vec{k}} \vec{n})
\Phi_{\vec{k}} \right)^2} 
\,\, \longrightarrow \,\,\text{min} \label{min1}
\end{eqnarray}
The physical resistivity is given by the minimum of
$\rho[\Phi_{\vec{k}}]$ regarded as a functional of $\Phi_{\vec{k}}$,
the scattering matrix $F$ is defined in (\ref{F}). Instead of solving
the Boltzmann equation exactly, we can guess the qualitative structure
of the solution: the strong scattering will equilibrate the
distribution function in a region of width $\Delta k$ around the hot
lines (see Fig.~\ref{phasediaHot.fig}). In the dirty limit, the
distribution function is of the well known form $\Phi_{\vec{k}}^0 =
\vec{v_K} \vec{E}$ and as a variational ansatz we assume that
$\Phi_{\vec{k}}$ vanishes (or more precisely is equal to $1/2
(\Phi_{\vec{k}}^0+\Phi_{\vec{k}\pm \Q}^0)$) in a region of width
$\Delta k$ around the hot lines.  It is easy to calculate the
qualitative structure of $\Delta \rho$ as a function of the
variational parameter $\Delta k$ in the regime where the disorder
dominates $\Delta \rho \lesssim \rho_0$:
\begin{eqnarray}
\Delta \rho[\Delta k] \approx x |\Delta k|
+ \frac{t^2}{(t+r+(\Delta k)^2)^{(4-d)/2}} +O(t^2) \label{var}
\end{eqnarray}
The first term is due to the denominator in (\ref{min1}) and reflects
the fact that the disorder favors a smooth distribution function. The
structure of the second term can easily be guessed (or calculated). It
reproduces the well-known results e.g. of Moriya\cite{moriya} for
$\Delta k=0$, who essentially averages the scattering rates $F_{\k
\k'}$ over the full Fermi surface.

In the dirty limit, the first term in (\ref{var}) dominates, $\Delta k$
vanishes at the minimum and one obtains the well-known 
$$\Delta \rho \approx \frac{t^2}{(t+r)^{(4-d)/2}}, 
\quad \text{for }
\Delta k \ll t+r.$$
In the ultra-clean limit ($x \ll t^2$), $\Delta k$ is of order $\kf$ and
$$\Delta \rho \approx t^2 \quad \text{for }
\Delta k \approx 1.$$
For clean systems there is an interesting intermediate regime where
$t+r<(\Delta k)^2<1$, where we obtain a non-trivial result by minimizing
(\ref{var})
$$ \Delta \rho \approx t^{\frac{2}{5-d}} x^{\frac{4-d}{5-d}} 
\quad \text{with }
\sqrt{t+r}<\Delta k \approx (t^2/x)^{\frac{1}{5-d}} < 1.$$
This very simple calculation reproduces the results of (\ref{scaleB0})
qualitatively.

\section{Relaxation time approximations} \label{relaxTime}

Strictly speaking, a relaxation time approximation is not valid near a
quantum critical point, where the scattering is dominated by processes
involving a fixed momentum transfer $\vec{Q}$.  The time to
equilibrate a quasiparticle distribution at the point $\vec{k}$ on the
Fermi surface will crucially depend on the quasiparticle distribution
at the point $\vec{k}+\vec{Q}$. For example if $\vec{Q}$ is
perpendicular to the electric field $\vec{E}$ and some reflection or
gliding plane (and if the Fermi surface is not too complicated), then
$\Q$ will connect points with the same non-equilibrium quasiparticle
distribution and the singular scattering will {\em not} relax the
non-equilibrium distribution. In such a situation vertex corrections
cancel the leading self-energy contributions and the relaxiation time
approximation is completely wrong (see also [\onlinecite{vertexHall}]
on the role of vertex corrections for the Hall effects).

Nevertheless, the relaxation time analysis gives qualitatively the
correct resistivity for generic situations, where the $\Q$-vectors are
not perpendicular to $\vec{E}$ as can be verified by comparison with
the full solution of the Boltzmann equations discussed in the main
text.  The relaxation time approximation can have much larger errors
for the Hall effect because the distribution function $\Phi_\k$
perpendicular to the current is often dominated by quasiparticles
which are scattered into this region by spin-fluctuations with
momentum $\Q_i$ (in the geometry show in Fig.~\ref{phasediaHot.fig}
this effect is absent).

In second order perturbation theory the life-time of the electron at
$T=0$ is given by \cite{hlubina} \begin{eqnarray}
\frac{1}{\tau_{\vec{k}}}= 2 g_S^2/\Gamma \sum_{\vec{k}'}
\int_0^{\epsilon_{\vec{k}}}\! \!\!  d \w \IM
\chi_{\vec{k}-\vec{k}'}(\w)
\delta(\w-(\epsilon_{\vec{k}}-\epsilon_{\vec{k}'})), \end{eqnarray}
where $\chi$ is given by (\ref{chi}) with $\w_{\vec{q}}=(\vec{q}\pm
\Q_i)^2/q_0^2$. Splitting the integration over $\vec{k'}$ in a surface
integral over the Fermi surface and an energy integral perpendicular
to it $\int d^3 \vec{k}'= \surfint d \vec{k}`/v_F \int d
\epsilon_{\vec{k}'}$, where $v_F(\vec{k}')$ is the Fermi
velocity, performing first the $\epsilon_{\vec{k}'}$ and then the
$\w$ integration we obtain 
\begin{eqnarray}
\frac{1}{\tau_{\vec{k}}}&\approx& \frac{g_S^2 }{v_F (2 \pi)^3} \surfint
d\vec{k}' \ln \left[
\frac{(\epsilon_{\vec{k}}/\Gamma)^2+(r+\w_{\vec{k}-\vec{k}`})^2}
{(r+\w_{\vec{k}-\vec{k}`})^2}\right]
\\ &\approx& \frac{g_S^2 q_0^2}{v_F (2 \pi)^2}
\frac{\epsilon_{\vec{k}}}{\Gamma} \text{min}\!\left[
\frac{\epsilon_{\vec{k}}}{2 \Gamma \delta_k^2},1 \right], \quad
\delta_k^2=r+(\delta \vec{k}/q_0)^2 \end{eqnarray} 
where $\delta
\vec{k}$ is the distance of  $\vec{k}+\vec{Q}$ from the Fermi surface,
i.e. essentially the distance of $\vec{k}$ from  lines on the Fermi surface,
where $\epsilon_\k=\epsilon_{\k\pm \Q_i}=\mu$. Along these ``hot lines''
the inelastic scattering is strongest.

The scattering rate depends crucially on the distance from the hot
lines $\delta \kappa=\delta k/\kf \approx \delta k/q_0$.  Far from the
hot lines or at lowest energies for a system not directly at the
quantum critical point, we recover the usual $1/\tau_{\vec{k}} \propto
\epsilon_{\vec{k}}^2$. Directly at the QCP and for $\delta \kappa=0$
 the scattering rate is linear in the quasiparticle energy. At
finite temperatures a typical quasi particle has the energy $T$ and
for qualitative estimates we use the approximation
\begin{eqnarray}
\frac{1}{\tau_{\vec{k}}^s} &\approx& 
\frac{1}{\tau_M} \left\{ \begin{array}{ll} t^2/(r+t) & \text{for }
\delta \kappa < \sqrt{r+t}\\ t^2/(\delta \kappa)^2 & \text{for }
\delta \kappa > \sqrt{r+t} \end{array}\right. \label{tauTot}
\end{eqnarray} where $\tau_M$ is a typical scattering rate at the
temperature $\Gamma$ and $t=T/\Gamma$ the dimensionless
temperature. For $r\ll 1$ the scattering is highly anisotropic and
strongest in a region of width $\delta \kappa \approx \sqrt{r+t}$. In
the presence of weak disorder, the elastic scattering due to
impurities has to be added \begin{eqnarray}\label{tauT}
\frac{1}{\tau_{\vec{k}}}=\frac{1}{\tau_{\text{el}}}+
\frac{1}{\tau_{\vec{k}}^s} \approx x
\frac{1}{\tau_M}+\frac{1}{\tau_{\vec{k}}^s} \end{eqnarray} where me
measure the relative strength of impurity scattering by the
dimensionless quantity $x\approx \tau_M/\tau_{\text{el}}\approx
1/{\text{RRR}}$ which is defined in (\ref{defConst}).

The conductivity in relaxation time approximation is proportional
to the average of $\tau_{\vec{k}}$ over the Fermi surface (assuming
that the Fermi velocity does not vary too strongly).
By integrating  (\ref{tauTot}) over $\delta \kappa$, we therefore obtain
an estimate for the conductivity.
The conductivity $\sigma$ in
units of the conductivity $\sigma_M=1/\rho_M$
is approximately given by 
\begin{eqnarray}\label{sigmaT1}
\frac{\sigma(T)}{\sigma_M} &\approx& 
\left\langle  \tau_{\vec{k}} \right\rangle/\tau_M =
\left\langle  \left(\frac{\tau_M}{\tau_{\text{el}}}+
\frac{\tau_M}{\tau_{\vec{k}}^s}\right)^{-1} \right\rangle  \\
&\approx& 
\int_0^{\sqrt{r+t}}\! \! \! d \delta \kappa \frac{1}{x+t^2/(r+t)} +
\int_{\sqrt{r+t}}^1 \! \! \! d \delta \kappa 
\frac{1}{x+t^2/(\delta \kappa)^2}  \nonumber
\\
&=&\frac{1}{x}\biggl(1-\frac{t^2/\sqrt{r+t}}{x+t^2/(r+t)} \label{sigmaTA}\\
&& \hphantom{\frac{1}{x}\biggl()}
-\frac{t}{\sqrt{x}}
\left(  \arctan \frac{\sqrt{x}}{t} 
-\arctan \frac{\sqrt{x} \sqrt{r+t}}{t}\right)\biggr) \nonumber
\end{eqnarray}

In the ultra-clean limit, $x\ll t^2<1$, the main contribution arises from
regions far from the hot line or large $\delta \kappa$. The conductivity
diverges with $1/T^2$ with a prefactor of order $1$
\begin{eqnarray}
\rho(\Gamma t) \approx t^2 \text{ for } 
\frac{1}{\tau_{\text{el}}} \ll \min \! 
\left[ \frac{1}{\tau_{\vec{k}}^s}\right] \text{ or }  x \ll t^2  .
\end{eqnarray}
as has been pointed out by Hlubina and Rice \cite{hlubina}.
In the dirty limit, $x \gg t^2/(r+t)$, the elastic scattering
dominates and one can expand (\ref{sigmaT1}) or (\ref{sigmaTA}) in the
inelastic scattering rates:
\begin{eqnarray}
\frac{\Delta \rho(T)}{\rho_M} &\approx& \tau_M \left\langle
\frac{1}{\tau_{\vec{k}}^s}\right\rangle \approx 
2 \frac{t^2}{\sqrt{r+t}}  \\
&& \
\text{ for } 
\frac{1}{\tau_{\text{el}}} \gg \max \! 
\left[ \frac{1}{\tau_{\vec{k}}^s}\right] 
\text{ or }  t \ll \max [\sqrt{r x},x]
\nonumber
\end{eqnarray}
and we recover the well known $\delta \rho \propto T^{3/2}$ near the
quantum-critical point.

In the relaxation time approximation we obtain similar to our results
for the Boltzmann equation a pronounced intermediate regime, where
$\min[ 1/\tau_{\vec{k}}^s] \lesssim 1/\tau_{\text{el}} \lesssim
\max[1/\tau_{\vec{k}}^s]$. This corresponds to the regime where the
first $\arctan$ in (\ref{sigmaTA}) is $\pi/2$, while the second is
small, i.e.  $\max[\sqrt{r x},x ]<t<\sqrt{x}$. The rise of the
resistivity in {\em linear} in temperature in this regime:
\begin{eqnarray}
\frac{\Delta \rho(T)}{\rho_M} \approx t  \frac{\pi}{2} \sqrt{x} \text{ for }
\max[\sqrt{r
x},x] <  t <  \sqrt{x} . \nonumber
\end{eqnarray}

\section{Numerical methods for solving the Boltzmann equation}
The Boltzmann equations were solved by direct matrix inversion.  The
Hall effect and the $B^2$-magnetoresistivity were calculated in
perturbation theory in $B$. To keep the size of the matrices small it
is important to use all the symmetries of the problem and to choose a
suitable discretization of the Fermi surface. At low temperatures the
distribution function $\Phi_{\k}$ changes very rapidly perpendicular
to the hot lines. Therefore we define the distribution function on
polygon-shaped patches on the Fermi surface which are relatively long
parallel to the hot lines $\Delta k_\| \ll k_F$ but short in the
perpendicular direction close to the hot lines $\Delta k \ll k_F
\min[\sqrt{t},t/\sqrt{x}]$. Because the kernel of the integral
equation is strongly peaked it is essential to calculate a small
number of matrix elements $M_{i j}=\surfint_{\text{patch } i} dk
\surfint_{\text{patch } j} dk' G_{k k'}$ numerically if they involve a
momentum transfers $\k-\k'$ close to $\Q_i$.  To speed up the
calculations one can approximate $I[y]\approx \pi^2 (2 \pi+y)/( y (4
\pi^2 + 6 \pi y+ 3 y^2))$ which is asymptotically exact in
next-to-leading order for large and small arguments; errors are
smaller than $1.5\%$ (The previously used\cite{hlubina,roschPRL}
approximation $I[y]\approx \pi^2/(y(3 y+2 \pi)$ is also sufficient to
reproduce next to all of our results with essentially neglegible
error, one exception is the temperature dependence of the Hall effect
in the ultra-clean limit, where it wrongly predicts $R_H \sim c_1 \pm
c_2 T$ instead of the correct $R_H \sim c_1 \pm c_2 T^{3/2}$).  We
used Fermi surfaces with cubic symmetry. As the largest irreducible
representation of the 48-dimensional cubic group is only three
dimensional, it was possible to reduce the size of the matrices by the
huge factor $16 \times 16$ using the full symmetry of the scattering
matrix. We didn't try to solve the Boltzmann equations numerically in
a finite magnetic field, which would break the cubic symmetry, but
used perturbation theory in $B$ to calculate the figures of this
paper.

It is essential to use a proper discretization of the magnetic field term
which conserves
\begin{eqnarray}\label{midB}
&& \surfint_S [\vec{B} \cdot (\vec{v}_\k \times \vec{\nabla}_\k \Phi_\k)] 
(d^2 \k/v_\k) \\
&& = \surfint_S (\vec{d}^2 \k
 \times \vec{\nabla}_\k \Phi_\k)  \cdot \vec{B} = \oint_{\partial S} \Phi_\k 
(\vec{B} \cdot d \k) \nonumber
\end{eqnarray}
 for an arbitrary part $S$ of the Fermi surface. Otherwise the
discretization leads to large errors. To determine the averaged
$\vec{v}\times \vec{B}$ contribution for a single patch, (\ref{midB})
is applied for a single patch. $\Phi_\k$ on the edges of the patch $i$
is determined by a linear interpolation: $ [\vec{B} \cdot (\vec{v}_\k
\times \vec{\nabla}_\k \Phi_\k)] (d^2 \k/v_\k) \approx \sum_j (\vec{B}
\cdot \Delta \k^i_{j})(
\bar{\Phi}_{\k_{i,j,2}}+\bar{\Phi}_{\k_{i,j,2}})/2$ where $\Delta
\k^i_{j}=\k_{i,j,2}-\k_{i,j,1}$ describes the $j$th edge with corners
$\k_{i,j,2}$ and $\k_{i,j,1}$. The vectors $\Delta \k^i_{j}$ point
counterclockwise around patch $i$ with $\sum_j \Delta \k^i_{j}=0$.
The value of $\bar{\Phi}_\k$ at the corners $\k_{i,j,n}$ ($n=1,2$) is
determined by a simple average over the value of $\Phi$ on each
neighboring patch.  The details of these averaging procedures are not
so important as long as (\ref{midB}) is not violated.

\section{Models used for the numerical calculation}
\label{appendixModelA}
For our numerical calculations we use three different models.  Model A
is based on a spherical Fermi-surface (see Fig.~\ref{phasediaHot.fig})
and the ordering wave-vector is $\vec{Q}=(0,0,\pm 2 \kf \cos \Theta_H$
with $\Theta_H =\pi/6$. The precise value of $\theta_H$ are not very
important as long as one stays away from $\theta_H=0$ (``$2 \kf$''
ordering) or $\theta_H=\pi/2$ (ferromagnetic ordering) where our
approach breaks down \cite{millis}; in our numerical calculations we
use $\theta_H=\pi/6$.  The dispersion $\w_{\vec{q}}$ in (\ref{chi}) is
given by $(\vec{q}-\vec{Q})^2/q_0^2$.

Model B describes a cubic system with commensurate spin waves with
ordering vectors $\vec{Q}=(\pm 1,\pm 1,\pm 1)$ and $\w_{\vec{q}}=3 +
\sum_{i=x,y,z} \cos \pi q_i$. The bandstructure was choosen to be
featureless but it includes terms which mix the various directions and
it is characterized by 8 non-intersecting ``hot lines'': $\epsilon_{\vec{k}}
=t \sum_{x,y,z} \cos \pi k_i+ t' \sum_{x,y,z} \cos 2 \pi k_i+ t''
(\cos \pi k_x \cos \pi k_y+\cos \pi k_x \cos \pi k_z+\cos \pi k_y \cos
\pi k_z)-\mu$ with $t=-0.2, t'=0.05, t''=-0.045$ and $\mu=-0.0225$

We have compared model B to model C, where the ordering vector is
parallel to the principal axes: $\vec{Q}_i=(\pm 1,0,0), (0,\pm 1,0),
(0,0,\pm 1)$ using also a different band-structure. All results are
very similar compared to those of model B and are therefore not shown
in this paper.


\end{appendix}


\begin{references}
\bibitem{2dim} We do not consider situations as in CeCu$_{6-x}$Au$_x$
\cite{stockert} where the spin fluctuations are extremely anisotropic.
\bibitem{hotLifetime} In three dimensions the scattering rate of the
quasiparticles at the "hot lines" vanishes linear in $T$, therefore
the quasiparticle approximation is only "marginally" valid. The
situation is less clear in $d=2$, where the scattering rate at ''hot
spots'' \cite{hlubina} is proportional to $\sqrt{T}$.
\bibitem{alphaH1} For simple enough
 band structures in crystals with  mirror planes perpendicular to the
 ordering wave vectors $\Q_i$, the points $\k_H$ on the hot lines can
be mapped on $\k_H\pm \Q_i$ by a mirror transformation,
$v_{\k_H}=v_{\k_H\pm \Q_i}$ and $\alpha_H$ is $1$
\bibitem{defProblem} $\k\pm \Q_i$ is projected on the
Fermi surface, $\pm \Q_i$ is choosen in such a way
that $\k\pm \Q_i$ is close to the Fermi surface. These details are not
important, because $p_\k$ vanishes far from the hot lines. We also neglect
points on the Fermi surface where two hot lines cross, their contribution
is small.
\end{references}
\end{document}